\documentclass[aps,prx,twocolumn,superscriptaddress,showpacs,floatfix]{revtex4-2}
\usepackage{amsmath,amssymb,graphicx}
\usepackage{wasysym}
\usepackage[utf8]{inputenc}
\usepackage[T1]{fontenc}
\usepackage{xcolor}
\allowdisplaybreaks
\usepackage{rotating} 

\usepackage{textcomp}

\usepackage{bm}

\IfFileExists{siunitx.sty}{\usepackage{booktabs,siunitx}}{}

\usepackage{color}
\definecolor{LinkColor}{rgb}{0.256,0.439,0.588}
\usepackage{hyperref}

\usepackage{pifont}

\newcommand{\La}{\line (1,0  ){12}}
\newcommand{\Lb}{\line (0,10 ){11}}

\newcommand{\Ld}{\line (-1,0){12}}
\newcommand{\Le}{\line (0,-10){12}}

\newcommand{\C} {\circle*{4}}

\newcommand{\LaT}{\rule[-1pt]{0.4cm}{0.2em}}  
\newcommand{\LdT}{\rule[-1pt]{0.4cm}{0.2em}}  
\newcommand{\LbT}{\rotatebox{90}{\rule[-1pt]{0.4cm}{0.2em}}}  
\newcommand{\LeT}{\rotatebox{90}{\rule[-1pt]{0.4cm}{0.2em}}}  

\newcommand{\pA}{\put(-3,-10)}
\newcommand{\pB}{\put(9,-10)}
\newcommand{\pC}{\put(9,0)}

\newcommand{\pZ}{\put(-3,0)}

\newcommand{\pAT}{\put(-4,-10)} 
\newcommand{\pBT}{\put(8.2,-10)}  

\newcommand{\rhomb}{
  \pA{\C}\pB{\C}\pZ{\C}\pC{\C}
 }

\newcommand{\rhombH}{
  \begin{picture}(22,10)(-8,-7.8)
    \pA{\LaT}\pB{\Lb}\pZ{\Le}\pZ{\LdT}
    \rhomb
  \end{picture}
}
\newcommand{\rhombV}{
  \begin{picture}(22,10)(-8,-7.8)
   \pA{\La}\pBT{\LbT}\pAT{\LeT}\pC{\Ld}
    \rhomb
  \end{picture}
}

\usepackage{tikz}
\newcommand{\tikscale}{8.2pt}
\newcommand{\loopzig}{
\begin{tikzpicture}[baseline=1pt]
    \fill (0,0) circle (\tikscale/5);
    \fill (0,\tikscale) circle (\tikscale/5);
    \draw[line width=\tikscale/5] (0,0) -- (0*\tikscale,\tikscale);
    \fill (\tikscale,0) circle (\tikscale/5);
    \fill (\tikscale,\tikscale) circle (\tikscale/5);
    \draw[line width=\tikscale/5] (\tikscale,0) -- (1*\tikscale,\tikscale);
    \fill (2*\tikscale,0) circle (\tikscale/5);
    \fill (2*\tikscale,\tikscale) circle (\tikscale/5);
    \draw[line width=\tikscale/5] (2*\tikscale,0) -- (2*\tikscale,\tikscale);
    \fill (3*\tikscale,0) circle (\tikscale/5);
    \fill (3*\tikscale,\tikscale) circle (\tikscale/5);
    \draw[line width=\tikscale/5] (3*\tikscale,0) -- (3*\tikscale,\tikscale);
    \fill (4*\tikscale,0) circle (\tikscale/5);
    \fill (4*\tikscale,\tikscale) circle (\tikscale/5);
    \draw[line width=\tikscale/5] (4*\tikscale,0) -- (4*\tikscale,\tikscale);
    \draw[line width=\tikscale/5] (0*\tikscale,0) -- (0*\tikscale+\tikscale,0);
    \draw[line width=\tikscale/5] (0*\tikscale+\tikscale,\tikscale) -- (0*\tikscale+2*\tikscale,\tikscale);
    \draw[line width=\tikscale/5] (2*\tikscale,0) -- (2*\tikscale+\tikscale,0);
    \draw[line width=\tikscale/5] (2*\tikscale+\tikscale,\tikscale) -- (2*\tikscale+2*\tikscale,\tikscale);
    \draw[line width=0.4pt] (0*\tikscale,\tikscale) -- (0*\tikscale+\tikscale,\tikscale);
    \draw[line width=0.4pt] (0*\tikscale+\tikscale,0) -- (0*\tikscale+2*\tikscale,0);
    \draw[line width=0.4pt] (2*\tikscale,\tikscale) -- (2*\tikscale+\tikscale,\tikscale);
    \draw[line width=0.4pt] (2*\tikscale+\tikscale,0) -- (2*\tikscale+2*\tikscale,0);
    \draw[line width=\tikscale/5] (0,\tikscale) -- (-\tikscale*0.5,\tikscale);
    \draw[line width=0.4pt] (0,0) -- (-\tikscale*0.5,0);
    \draw[line width=\tikscale/5] (4*\tikscale,0) -- (\tikscale*4.5,0);
    \draw[line width=\tikscale/5] (0,\tikscale) -- (-\tikscale*0.5,\tikscale);
    \draw[line width=\tikscale/5] (4*\tikscale,0) -- (\tikscale*4.5,0);
    \draw[line width=0.4pt] (4*\tikscale,\tikscale) -- (\tikscale*4.5,\tikscale);
    \fill (4*\tikscale+2+1*\tikscale/2,\tikscale/2) circle (\tikscale/10);
    \fill (-1*\tikscale/2-2,\tikscale/2) circle (\tikscale/10);
    \fill (4*\tikscale+2+2*\tikscale/2,\tikscale/2) circle (\tikscale/10);
    \fill (-2*\tikscale/2-2,\tikscale/2) circle (\tikscale/10);
    \fill (4*\tikscale+2+3*\tikscale/2,\tikscale/2) circle (\tikscale/10);
    \fill (-3*\tikscale/2-2,\tikscale/2) circle (\tikscale/10);    
\end{tikzpicture}
}
\newcommand{\loopparallel}{
\begin{tikzpicture}[baseline=1pt]
    \fill (0,0) circle (\tikscale/5);
    \fill (0,\tikscale) circle (\tikscale/5);
    \draw[line width=0.4pt] (0,0) -- (0*\tikscale,\tikscale);
    \fill (\tikscale,0) circle (\tikscale/5);
    \fill (\tikscale,\tikscale) circle (\tikscale/5);
    \draw[line width=0.4pt] (\tikscale,0) -- (1*\tikscale,\tikscale);
    \fill (2*\tikscale,0) circle (\tikscale/5);
    \fill (2*\tikscale,\tikscale) circle (\tikscale/5);
    \draw[line width=0.4pt] (2*\tikscale,0) -- (2*\tikscale,\tikscale);
    \fill (3*\tikscale,0) circle (\tikscale/5);
    \fill (3*\tikscale,\tikscale) circle (\tikscale/5);
    \draw[line width=0.4pt] (3*\tikscale,0) -- (3*\tikscale,\tikscale);
    \fill (4*\tikscale,0) circle (\tikscale/5);
    \fill (4*\tikscale,\tikscale) circle (\tikscale/5);
    \draw[line width=0.4pt] (4*\tikscale,0) -- (4*\tikscale,\tikscale);
    \draw[line width=\tikscale/5] (0*\tikscale,0) -- (0*\tikscale+\tikscale,0);
    \draw[line width=\tikscale/5] (0*\tikscale+\tikscale,\tikscale) -- (0*\tikscale+2*\tikscale,\tikscale);
    \draw[line width=\tikscale/5] (2*\tikscale,0) -- (2*\tikscale+\tikscale,0);
    \draw[line width=\tikscale/5] (2*\tikscale+\tikscale,\tikscale) -- (2*\tikscale+2*\tikscale,\tikscale);
    \draw[line width=\tikscale/5] (0*\tikscale,\tikscale) -- (0*\tikscale+\tikscale,\tikscale);
    \draw[line width=\tikscale/5] (0*\tikscale+\tikscale,0) -- (0*\tikscale+2*\tikscale,0);
    \draw[line width=\tikscale/5] (2*\tikscale,\tikscale) -- (2*\tikscale+\tikscale,\tikscale);
    \draw[line width=\tikscale/5] (2*\tikscale+\tikscale,0) -- (2*\tikscale+2*\tikscale,0);
    \draw[line width=\tikscale/5] (0,\tikscale) -- (-\tikscale*0.5,\tikscale);
    \draw[line width=\tikscale/5] (0,0) -- (-\tikscale*0.5,0);
    \draw[line width=\tikscale/5] (4*\tikscale,0) -- (\tikscale*4.5,0);
    \draw[line width=\tikscale/5] (0,\tikscale) -- (-\tikscale*0.5,\tikscale);
    \draw[line width=\tikscale/5] (4*\tikscale,0) -- (\tikscale*4.5,0);
    \draw[line width=\tikscale/5] (4*\tikscale,\tikscale) -- (\tikscale*4.5,\tikscale);
    \fill (4*\tikscale+2+1*\tikscale/2,\tikscale/2) circle (\tikscale/10);
    \fill (-1*\tikscale/2-2,\tikscale/2) circle (\tikscale/11);
    \fill (4*\tikscale+2+2*\tikscale/2,\tikscale/2) circle (\tikscale/11);
    \fill (-2*\tikscale/2-2,\tikscale/2) circle (\tikscale/11);
    \fill (4*\tikscale+2+3*\tikscale/2,\tikscale/2) circle (\tikscale/11);
    \fill (-3*\tikscale/2-2,\tikscale/2) circle (\tikscale/11);    
\end{tikzpicture}
}

\newcommand{\bea}{\begin{eqnarray}}
\newcommand{\eea}{\end{eqnarray}}

\newcommand{\vdimer}{{\vrule height0.2cm width0.05cm depth0pt}}
\newcommand{\hdimer}{{\hrule height0.05cm width0.2cm depth0pt}}

\newcommand{\mdimer}{\vbox{\hdimer \vskip 0.0625cm}}

\newcommand{\overdimer}{\hbox{\hskip 0.05cm \vdimer }}
\newcommand{\ohordimer}{\hbox{\vbox{\mdimer}}}

\begin{document}

\title{Classical fully packed loop model with attractive interactions on the square lattice}

\author{Bhupen Dabholkar}
\thanks{These authors contributed equally to this work.}
\affiliation{Laboratoire de Physique Th\'eorique, Universit\'e de Toulouse, CNRS, UPS, France}

\author{Xiaoxue Ran}
\thanks{These authors contributed equally to this work.}
\affiliation{Department of Physics and HKU-UCAS Joint Institute of Theoretical and Computational Physics,The University of Hong Kong, Pokfulam Road, Hong Kong SAR, China}

\author{Junchen Rong}
\affiliation{Institut des Hautes \'Etudes Scientifiques, 91440 Bures-sur-Yvette, France}

\author{Zheng Yan}
\affiliation{Department of Physics and HKU-UCAS Joint Institute of Theoretical and Computational Physics,The University of Hong Kong, Pokfulam Road, Hong Kong SAR, China}

\author{G. J. Sreejith}
\affiliation{IISER Pune, Dr Homi Bhabha Road, Pune 411008, India}

\author{Zi Yang Meng}
\email{zymeng@hku.hk}
\affiliation{Department of Physics and HKU-UCAS Joint Institute of Theoretical and Computational Physics,The University of Hong Kong, Pokfulam Road, Hong Kong SAR, China}

\author{ Fabien Alet}
\email{fabien.alet@cnrs.fr}
\affiliation{Laboratoire de Physique Th\'eorique, Universit\'e de Toulouse, CNRS, UPS, France}

\begin{abstract}

We study a classical model of fully packed loops on the square lattice, which interact through attractive loop segment interactions between opposite sides of plaquettes. This study is motivated by effective models of interacting quantum matter arising in frustrated magnets or Rydberg atom arrays, for which loop degrees of freedom appear at low energy. Through the combination of Monte Carlo simulations and of an effective height field theory, we find that the critical point known to occur at infinite temperature gives rise to a high-temperature critical phase with floating exponents. At lower temperature, the system transitions via a Kosterlitz-Thouless phase transition to a nematic phase where lattice rotation symmetry is broken. We discuss consequences for the phase diagram of the quantum loop model on the same lattice. 
\end{abstract}

\date{\today}
\maketitle

\section{Introduction}

An important notion in the renormalization group theory is the emergence of effective degrees of freedom at low energies. These new degrees of freedom can have local structures which take the form of a {\it constraint}. For instance, for degrees of freedom that live on the bonds of a lattice, a gauge-like condition can emerge which requires that every site of the lattice is touched by a fixed number of occupied bonds. 
Related statistical mechanical models such as dimer or loop models arise as effective theories in many physical situations, such as in frustrated magnetic systems~\cite{moessnerResonating2001,moessnerIsing2001} as, {\it e.g.} in spin ice~\cite{bramwell}, Rydberg atom arrays~\cite{semeghiniProbing2021,rhineQuantum2021,ebadiQuantum2021,yanTriangular2022}, models of high-${\rm T}_c$ superconductors~\cite{rokhsarSuperconductivity1988}, adsorption physics~\cite{blunt08}, quantum Hall effects~\cite{gruzberg99,read01}, topological order~\cite{fendley08}, deconfined quantum critical points~\cite{Nahum1,nahum15,AletUnconventional,PowellContinuumTheory,AletGaugeTheory08,PowellClassical2Quantum,GangChen3DDimer,SreejithScalingDim,SreejithSO5,SreejithMonopoleFugacity}, etc. Loop models also have a long history in statistical physics~\cite{nienhuis2010,nienhuis87,jacobsen09,Nahum1,nahum13}, in relation to Potts models~\cite{Kondev1996}, Temperley-Lieb algebras~\cite{temperley71}, polymers and O($N$) models~\cite{de_gennes72,jacobsen03}, Schramm-Loewner evolution~\cite{cardy05}, or percolation.
These models often assign fixed fugacity for loops~\cite{nienhuis87,nienhuis2010}, but there are few results when the loop segments interact~\cite{jacobsen09b}, even though loop interactions naturally arise in effective models of quantum condensed matter~\cite{rokhsarSuperconductivity1988,schwandt2010}.

In this work, we study a two-dimensional (2D) classical statistical mechanical model of fully packed loops which attract locally. With the help of a directed-loop Monte Carlo algorithm~\cite{barkema1998monte,alet2006classical,sandvik2006correlations,syljuasenquantum2002,syljuaasen2004directed,aletdirected2003} and a Coulomb gas~\cite{nienhuis87} approach formulated in terms of a height-field description of the loop constraint~\cite{kondev96b,moessner2004}, we obtain evidence for the existence of a finite-temperature Kosterlitz-Thouless (KT) transition separating a high-temperature critical phase from a low-temperature nematic phase. Our results have similarities with those obtained for the classical dimer model with attractive interactions~\cite{alet2005Interacting,alet2006classical,papanikolaou2007quantum}, albeit with specific differences that we highlight.

Aside from their interest in two-dimensional statistical mechanics in extending previous works on loop models~\cite{nienhuis2010,nienhuis87,jacobsen03,jacobsen09,nahum13,kundu2023flux}, our results are also relevant for quantum-constrained models. First, the ground-state wave function at a Rokhsar-Kivelson point~\cite{rokhsarSuperconductivity1988} (or its generalizations~\cite{castelnovo2005,Balasubramanian22}) in the phase diagram of quantum loop models (QLM) maps to the partition function of a classical loop model. It is possible to construct extended quantum loop models (following the prescription in Ref.~\cite{castelnovo2005}, see an example in Ref.~\cite{castelnovo2006} for a dimer model) whose ground-phase diagram is entirely given by the finite-temperature phase diagram of a classical (interacting) loop model, as the one we describe in this work. The second connection is made by realizing that the phase diagram of the classical model and the methods we use in its inference can serve to guide us in mapping out the finite-temperature phase diagram~\cite{henry14} and transitions of the quantum loop model~\cite{shannonCyclic2004,henry14,syljuasen06,Plat2015Magnetization,Roychowdhury2015topological,ranFully2022,yanFully2022} (see e.g. the finite-temperature phase diagram of the quantum dimer model~\cite{dabholkar2022reentrance}). Such quantum-constrained models host a rich set of phases~\cite{yanWidely2021,yanHeight2022,Verresen22,yanTriangular2022,yanTopological2020,Pollmann2011,Banerjee2013,ranFully2022,yanFully2022,Plat2015Magnetization,Roychowdhury2015topological,shannonCyclic2004,henry14,syljuasen06} and have recently been shown to be relevant in the context of Rydberg atom arrays~\cite{Browaeys20,Glaetzle14,Celi20,Verresen21,semeghiniProbing2021,rhineQuantum2021,yanTriangular2022,yanEmergent2023}, where the Rydberg blockade effectively implements the loop or dimer constraint. 

The rest of the paper is organized as follows: In Sec.~\ref{sec:ii}, we introduce the classical loop model. Section~\ref{sec:iii} introduces the different physical observables computed in this work. Section~\ref{sec:iv} provides a field theoretical perspective to the model and phase diagram in the form of a Coulomb gas analysis. The Monte Carlo simulation results and their analysis are given in Sec.~\ref{sec:v}, in which, Sec.~\ref{sec:va} presents winding number fluctuations and Sec.~\ref{sec:vb} an analysis of the low-temperature order parameter and its susceptibility. In Sec.~\ref{sec:vc}, we discuss the behavior of various correlation functions in the high-temperature phase. These results are analyzed in light of the Coulomb gas predictions of Sec.~\ref{sec:iv}. We present our  conclusions and some perspectives in Sec.~\ref{sec:vi}. Appendix~\ref{sec:appI} describes the Monte Carlo directed loop algorithm used in our numerical study and Appendix~\ref{sec:appII} contains further results on correlation functions.

\section{Model and methods}
\label{sec:ii}
\begin{figure}[htp!]
	\centering
	\includegraphics[width=1\columnwidth]{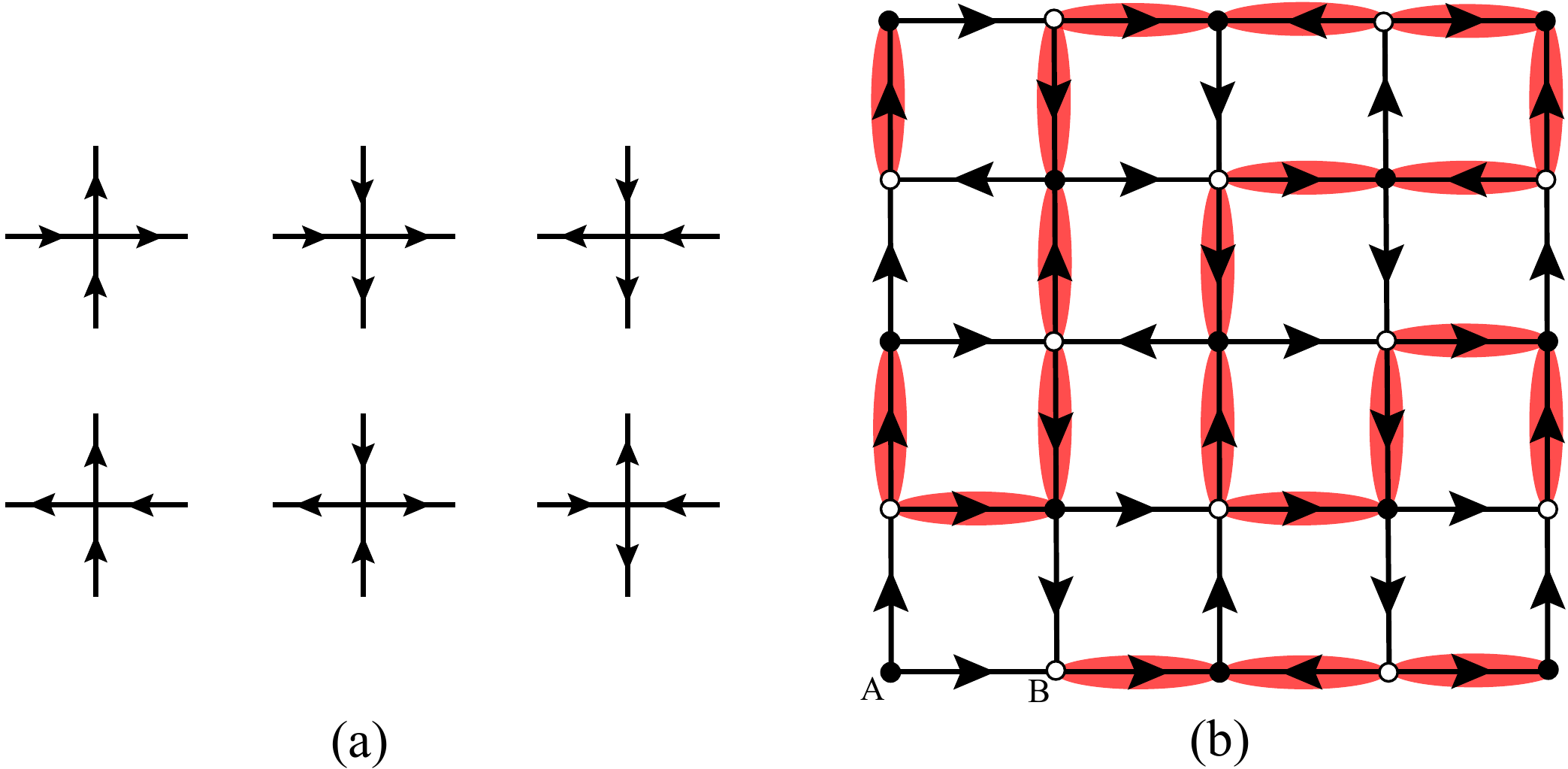}
	\caption{(a) The allowed vertex types for the 6-vertex model. (b) Correspondence between a 6-vertex configuration and the fully packed loop configuration on the square lattice. The solid and open circles represent sites of the A and B sublattices, respectively. Placing dimers on all incoming arrows of the vertices on the A-sublattice produces a fully-packed loop configuration.}
	\label{fig:six}
\end{figure}

{\it Configurations. --- } Configurations of the fully packed loop model on a square lattice require two loop segments (or ``dimers'') to touch each site of a square lattice, and are in one-to-one correspondence with configurations of the six-vertex model~\cite{LiebWu,Lieb1967Residual,Lieb1967Exact,Lieb1967Exact2,Sutherland1967Exact,Lieb1967Exact3}. The ice-rule constraint of the six-vertex model associates an arrow on each bond and only allows vertices which have two arrows pointing inwards and two outwards from the lattice site. Under this constraint, there are six possible vertex configurations on the square lattice as shown in Fig.~\ref{fig:six} (a). The mapping from the six-vertex model to the loop model on the square lattice 
is illustrated in Fig.~\ref{fig:six} (b). If we place dimers on two incoming arrows on all sites of a sublattice of the square lattice, dimers will collectively form fully packed loops as every site is touched by exactly two dimers (`loop segments'). 

{\it Energetics --- } Loop or vertex models often associate a fugacity with each closed loop or to each type of vertex respectively, to define the corresponding partition function~\cite{nienhuis2010}. The model that we study here associates an interaction energy term between proximate parallel loop segments, similar to the classical interacting dimer models~\cite{alet2005Interacting,alet2006classical}. 
We consider the following  partition function and energy for an interacting fully-packed loop model on the square lattice  
\begin{eqnarray}\label{eq:eq1}
    Z &=& \sum_{c} e^{-\beta E_{c}}\nonumber \\
    E_{c} &=& V(N(\rhombV) + N(\rhombH)),
\end{eqnarray}
where the summation in the partition function $Z$ is over all fully-packed loop configurations on the square lattice and $\beta=1/T$ is the inverse temperature. We assign an energy $E_{c}$ to each covering in which $(N(\rhombV) + N(\rhombH))$ counts the number of plaquettes with parallel loop segments. Note that there is no energy assigned to a plaquette that has more than two loop segments. Here we set $V=-1$, which corresponds to {\it attractive} interactions between loop segments. We assume periodic boundary conditions for square lattices of linear size $L$ with $N=L^2$ total number of sites. The model Eq.~\eqref{eq:eq1} is the limiting case of the quantum loop model on the square lattice~\cite{shannonCyclic2004} obtained when kinetic terms vanish in the QLM. To the best of our knowledge, this fully-packed loop model with aligning interactions has never been studied earlier.

{\it Limiting cases --- } The model admits two simple limits. At infinite temperature, it is equivalent to the 6-vertex model at the ice-point with equal fugacities for all vertices in Fig. \ref{fig:six} (a) which is {\it critical} with power-law correlators (see the precise description below). At $T=0$, there are two configurations which minimize the energy ($E_0=-L^2$). These are {\it nematic} configurations with $L$ horizontal or vertical loops that wrap around the boundary. The $\pi/2$ lattice rotation symmetry is broken at $T=0$, and since this model admits only discrete energies - the first excited states have energies $E_1=-L^2+4$ - we expect a finite-temperature transition into a low-temperature nematic phase. As will be shown below, this transition is of Kosterlitz-Thouless (KT) type.

While the two limiting phases (critical and nematic) are easily identified, one cannot exclude other intervening phases. We will explore the finite-temperature phase diagram of the model using directed-loop Monte Carlo simulation~\cite{barkema1998monte,alet2006classical,sandvik2006correlations,syljuasenquantum2002,syljuaasen2004directed,aletdirected2003}, which allows for efficient non-local moves. The precise implementation we use is presented in Appendix~\ref{sec:appI}. The simulations are supplemented by a field-theoretical analysis in terms of a Coulomb gas description of the system (Sec.~\ref{sec:iv}).

\section{Physical Observables}
\label{sec:iii}
In this section, we describe the observables measured during the Monte Carlo simulations to characterize the phases and the transitions.

{\it Winding number fluctuations --- } 
 Fully-packed loop configurations on the square lattice can be associated with two integer winding numbers $W_x$ and $W_y$. To compute $W_y$ ($W_x$), draw a horizontal (vertical) line that cuts across $L$ lattice bonds oriented in the $y$ ($x$) direction. For a given configuration, we denote by $N_o$ and $N_e$ the number of loop segments on the odd and even bonds that cross this line.
The winding numbers are defined as $N_e-N_o$. Each winding number $W_x$ and $W_y$ vary between $-L$ and $L$, and there is at least one fully-packed loop configuration for any pair $(W_x,W_y)$ in this range. Note that the loop constraint ensures that the winding numbers calculated using different parallel lines are the same.

On account of translation symmetry, the equilibrium average values of $W_x, W_y$ vanish, but not their fluctuations
\begin{equation}
\langle W^2 \rangle =\frac{1}{2} \langle W^2_{x}+W^2_{y} \rangle,
\label{eq:eq2}
\end{equation}
 which have useful physical content and can easily be measured in Monte Carlo simulations ~\cite{alet2005Interacting,alet2006classical,dabholkar2022reentrance,Pollock1987Path,Role1998Henelius}. 
{\it Low-temperature order parameter --- }  We can identify the low-temperature phase through the rotational symmetry breaking, nematic order parameter ~\cite{leungcolumnar1996,alet2006classical,papanikolaou2007quantum,yanFully2022,ranFully2022}
\begin{equation}
D=\frac{1}{N}|N_{\overdimer} - N_{\ohordimer}|,
\label{eq:orderparameter}
\end{equation}
with $N_{\overdimer} = \sum_{{\bf r}}n_{\overdimer}({\bf r})$ and $N_{\ohordimer} = \sum_{{\bf r}}n_{\ohordimer}({\bf r})$, where $n_{\ohordimer}({\bf r})$ denotes a horizontal loop segment at the site ${\bf r}$. It is 1 if a loop segment occupies the edge between ${\bf r}$ and ${\bf r}+(1,0)$ and is 0 if the edge is empty. $n_{\overdimer}({\bf r})$ denotes a vertical loop segment at site at lattice site ${\bf r}$, and is 1 if a loop segment occupies the edge between ${\bf r}$ and (${\bf r}+(0,1)$). The order parameter $D$ is $1$ in the two nematic ground-states, and vanishes ($\langle D \rangle =0$) at infinite temperature. 

We also compute the associated susceptibility \cite{leungcolumnar1996,alet2006classical,papanikolaou2007quantum}:
\begin{equation}
\chi_D = N(\langle D^{2} \rangle-\langle D\rangle^{2})
\label{eq:eq4}
\end{equation}
and monitor its temperature dependence. As shown below, the divergence of $\chi_D$ allows us to determine the transition temperature and the form of the divergence can be further used to infer the nature of the transition ~\cite{paiva2004critical,chenFermi2021,costaPhonon2018,jiangMonte2022}.

{\it Loop-segment (dimer) correlators --- } We consider the connected correlation function ~\cite{younblood1980,sutherland68,falco2013,alet2006classical,moessner2004} between loop-segments separated by a vector ${\bf r}=(x,y)$: $C_{\alpha,\beta}({\bf r})=\langle n_\alpha({\bf 0})n_\beta({\bf r})\rangle-1/4$, where $\alpha,\beta$ can be ${\ohordimer},{\overdimer}$. The expectation value $\langle n_\alpha \rangle \langle n_\beta \rangle = 1/4$ has been subtracted to get the connected correlator. In the Monte Carlo simulations, we average over all possible initial positions ${\bf 0}$ of the first loop segment, as well as all equivalent pairs $\alpha,\beta$. For simplicity, we will focus on the lattice direction ${\bf r}=(x=r,0)$ and consider three types of loop-segment correlations, longitudinal, transverse, and crossed, respectively defined as:
\begin{eqnarray}
C^L(r) & = & \langle n_{\ohordimer} ({\bf 0}) n_{\ohordimer} (r,0) \rangle-1/4,
\label{eq:eq5} \\
C^T(r)& = & \langle n_{\overdimer} ({\bf 0})n_{\overdimer} (r,0) \rangle-1/4, 
\label{eq:eq6} \\
C^C(r)& = & \langle n_{\ohordimer} ({\bf 0})n_{\overdimer} (r,0) \rangle-1/4.
\label{eq:eq7}
\end{eqnarray}
We will also consider correlators associated to the nematic order parameter
\begin{multline}
\langle D_{\bf 0}D_{\bf r}\rangle =\langle (n_{\overdimer}(0)-n_{\ohordimer}(0)) (n_{\overdimer}({\bf r})-n_{\ohordimer}({\bf r})) \rangle\\
=C^L(r)+C^T(r) - 2C^C(r)
\label{eq:eq8}
\end{multline}
where in the last line we again focus on the direction ${\bf r}=(x=r,0)$.

{\it Monomer correlators --- }
We also measure the monomer-monomer correlator ~\cite{Fisher1963Statistical,krauth2003pocket}, which requires going beyond the definition of the fully-packed loop configurations space by allowing two test monomers -- sites touched by only one dimer -- while the rest of the sites are all touched by two loop segments. No monomer is included in the fully-packed loop model, and the Monte Carlo configurations generated by the directed loop algorithm, {\it once a directed loop is finished}, do not contain monomers. However, during the intermediate steps of the Monte Carlo process, the directed loop algorithm precisely samples the extended phase space with monomers, allowing to sample the monomer correlator defined below (see Ref.~\cite{alet2005Interacting,alet2006classical,sandvik2006correlations} and Appendix~\ref{sec:appI} for details). We define monomer correlation function
\begin{equation}
M({\bf r})=\langle m({\bf 0})m({\bf r})\rangle.
\label{eq:eq9}
\end{equation}
where the presence of a monomer at site ${\bf r}$ is denoted by $m({\bf r})=1$  and $m({\bf r})=0$ otherwise. The monomer correlation function $M({\bf r})$ is estimated as the fraction of such Monte Carlo samples of configurations with two monomers separated by ${\bf r}$ ~\cite{krauth2003pocket,sandvik2006correlations,papanikolaou2007quantum}.

\section{Theoretical framework}
\label{sec:iv}
 Before analyzing the results of the numerical simulations, we first describe the finite-temperature phase diagram of the model using a field theoretical analysis. Following its success in two-dimensional models of statistical mechanics~\cite{andrews84,pasquier87,warnaar92,blote93,blote94,kondev96,kondev96b,kondev94,moessner2004,alet2006classical,papanikolaou2007quantum,wilkins2020}, we use a Coulomb gas description~\cite{nienhuis87}, formulated in terms of a height field $h(p)$. This field lives on plaquettes $p$ of the lattice, and is defined (up to an irrelevant constant) in the following way: when turning clockwise around A-sublattice sites of the square lattice, the height increases (decreases) by $1/2$ (i.e. $h\rightarrow h \pm 1/2$) if one crosses a loop segment (an empty edge). 
 At the microscopic level, it can be shown that the value of the height field inside a small patch, can be changed by $1$ without changing the local loop segment configuration, simply by a change of configuration at far away points (similar argument as for the dimer model~\cite{alet2006classical}). To see this, consider a region encircled by a pair of loops separated by a plaquette ($\loopparallel$). Local changes to these loops that convert them into a single zig-zag loop ($\loopzig$) surrounding the region changes the heights by 1 (holding heights on the exterior fixed) everywhere inside the region and irrespective of the distance from these loops.

 This indicates that the physical action should be invariant under height shifts $h \rightarrow h \pm 1$. Promoting the height field to the continuum $h({\bf r)}$, we expect the effective action to be:
\begin{equation}
    S=\int d^2r\, [ g(T) \pi (\nabla h({\bf r}))^2 + v \cos ( 4 \pi h({\bf r}))].
    \label{eq:S}
\end{equation}
Such a free compact boson model can be used to describe several two-dimensional statistical physics models, including the Tomonaga–Luttinger liquid (through bosonization, see e.g \cite{senechal2004introduction}) and the XXZ spin chain (see e.g. \cite{lukyanov_long-distance_2003}). We briefly justify this action below, and discuss its validity alongside the numerical results in Sec.~\ref{sec:v}. 

This action is of the sine-Gordon type~\cite{jose77,amit80}.
Here $g(T)$ is the Coulomb gas coupling constant, which depends on microscopic details and on temperature $T$. At infinite temperature, we have $g(T=\infty)=1/3$ from exact results for the six-vertex model at the ice point.~\footnote{The ice-point corresponds to the XXZ spin chain at $\Delta = 1/2$ in notations where $\Delta>0$ corresponds to ferromagnetic interactions. The value of $g$ is given by $\Delta=\cos(\pi g)$, see e.g. Ref.~\cite{lukyanov_long-distance_2003,hikihara1998correlation}.}.~This action displays the competition between the first term ($g(T) \pi (\nabla h({\bf r}))^2$) which alone describes the critical phase (rough in the height language) to be encountered at high temperature, and the $v \cos ( 4 \pi h({\bf r}))$ ``vertex'' term whose minima corresponds to the two nematic configurations for which the average height is constant (flat configurations) and takes values $\bar{h}=\pm 1/4$. $v$ can also depend on temperature but its exact dependency is not relevant as long as it remains positive such that the two nematic configurations are always favored.

In the Coulomb gas language and given the periodicity of the height $h\rightarrow h+1$ in the microscopic configurations, the later vertex term can be identified with an electric charge $e=2$ operator. 
This term is irrelevant at infinite temperature where $g=1/3$, but becomes relevant when $g \geq g_c=1$ (a general electric charge $e$ operator reads $\exp( i 2 e \pi h)$, and has scaling dimension $e^2/(2g)$, and thus becomes relevant when $g\geq e^2/4$).
As interactions favor the flat nematic phases, we expect $g$ to increase (from its $g(T=\infty)=1/3$ value) as the temperature is lowered. 

The Coulomb gas analysis predicts a Kosterlitz-Thouless phase transition~\cite{KT73,Kosterlitz74,jose77,amit80} from a high-temperature critical phase to the low-temperature nematic phase, and furthermore provides predictions for several observables. First, the winding fluctuations can be related to the Coulomb gas constant~\cite{alet2006classical}:
\begin{equation}
\langle W^2 \rangle =\sum_{n\in \mathbb {Z}}n^{2}e^{-g \pi n^2}/\sum_{n\in \mathbb {Z}}e^{-g \pi n^2},
\label{eq:eq11}
\end{equation}
as used in Fig.~\ref{fig:2} below. This allows in particular to extract the Kosterlitz-Thouless transition temperature $T_{KT}$ at the predicted critical Coulomb gas constant $g_c=1$.

Next, the leading terms for the dimer/loop segment occupation operator {\it in the continuum} have been identified in Ref.~\cite{moessner2004} as:
\begin{widetext}
\begin{eqnarray}\label{dimerOPE}
    n_{\ohordimer} ({\bf r}=(x,y)) & = & \frac{1}{2}+(-)^{x+y+1} \nabla_y h - \frac{X}{2i} (\exp(2 i \pi h({\bf r})) - \exp(- 2 i \pi h({\bf r}))) \\
n_{\overdimer} ({\bf r}=(x,y)) & = & \frac{1}{2}+(-)^{x+y} \nabla_x h + \frac{X}{2i} (\exp(2 i \pi h({\bf r})) - \exp(- 2 i \pi h({\bf r})))
\end{eqnarray}
\end{widetext}

The loop segment occupation is thus composed of a gradient part and a vertex part. 
The vertex part of the loop segment operator can be expressed in harmonics of $2 \pi h$ (as $h\equiv h+1$) and microscopic $\pi/2$ rotations of the model give $h \rightarrow -h$ and $h \rightarrow h+1/2$. 
It can be identified with an electric charge $e=1$ in the Coulomb gas. 

Note that the overall sign in front of the gradient depends on the convention for the height (odd or even sublattice). The constant $X$ cannot be fixed easily and we need an external exact solution (see below) -- in fact, we expect it to be re-normalized, that is to change with temperature.
This gives the following predictions for the leading terms of the correlators defined in Eqs.~\eqref{eq:eq5}-\eqref{eq:eq7}
\begin{widetext}
\begin{eqnarray}   
C^{L}({\bf r}=(x,y))&=&\langle n_{\ohordimer}(0) n_{\ohordimer}({\bf r}) \rangle -1/4  =  (-)^{x+y} A \frac{x^2-y^2}{(x^2+y^2)^2}+\frac{B}{(x^2+y^2)^{1/2g}} \label{eq:eq12} \\
C^{T}({\bf r}=(x,y))&=&\langle n_{\overdimer}(0) n_{\overdimer}({\bf r}) \rangle -1/4  =  (-)^{x+y} A \frac{y^2-x^2}{(x^2+y^2)^2}+\frac{B}{(x^2+y^2)^{1/2g}} \label{eq:eq13} \\
C^{C}({\bf r}=(x,y))&=&\langle n_{\ohordimer}(0) n_{\overdimer}({\bf r}) \rangle -1/4  =  (-)^{x+y} A \frac{2 x y}{(x^2+y^2)^2}-\frac{B}{(x^2+y^2)^{1/2g}} \label{eq:eq14}
\end{eqnarray}
\end{widetext}
The coefficient $A=\frac{1}{4 g \pi^2}$ is fixed by the operator product expansion \eqref{dimerOPE} and the two-point correlation function $\langle h(x) h(y)\rangle$ known exactly for free compact boson conformal field theory \cite{ginsparg1988curiosities}.
We also have $B=X^2/2$, however, its dependence on $g$ is not universal. At $T=\infty$, exact expressions for the XXZ spin chain~\cite{lukyanov_long-distance_2003} give $B \simeq  0.01795$, see Table 1 in Ref.~\cite{lukyanov_long-distance_2003} (see also Ref.~\cite{falco2013}).

Finally, we note that on the lattice, a monomer creates a dislocation of $\pm 1$ in the height field. The prediction of the monomer correlator decaying as \begin{equation} M(r) \propto r^{-g} \end{equation} follows~\cite{alet2006classical,papanikolaou2007quantum} from the identification of the monomer operator with the $m=\pm 1$ magnetic charge operator (the sign depends on the sublattice) with a scaling dimension $g m^2/2$. 

This interpretation parallels the one for the interacting classical dimer model~\cite{alet2005Interacting,alet2006classical,papanikolaou2007quantum} with the following three minor (albeit important for numerics) distinctions: 
(i) The infinite temperature value of the Coulomb gas constant $g=1/3$ renders the vertex contribution (scaling as $r^{-3}$) {\it subleading} with respect to the dipolar contribution (scaling as $r^{-2}$), which explains why it is often not reported in the polarization fluctuations for the 6-vertex model~\cite{younblood1980}. For the dimer problem, we have $g(T=\infty)=1/2$ and both terms contribute equally to the $r^{-2}$ decay of the dimer correlators~\cite{falco2013}.
(ii) The critical value of the Coulomb gas constant at the critical point is $g_c=1$ (instead of $g_c=4$ for the dimer model), consistent with the lower degeneracy of the ground-states ($2$ nematic ground-states instead of $4$ columnar ground-states for the dimer model) and resulting in a larger value of the anomalous dimension of the low-temperature order parameter $\eta_D={1/g_c}=1$ for the loop model (see below) instead of $\eta_D=1/g_c=1/4$ for the dimer model at their respective KT transitions ~\cite{alet2005Interacting}.
(iii) Lastly, we found the critical value for the winding fluctuations  $\langle W^2\rangle$ is much {\it larger} for the loop model, which allows for a statistically meaningful measurement in the Monte Carlo simulations. The very small value of $\langle W^2\rangle $ for the dimer model does not allow for an accurate Monte Carlo determination of the critical point using the value of the winding number fluctuations.

\begin{figure}[htp!]
\centering
\includegraphics[width=1\columnwidth]{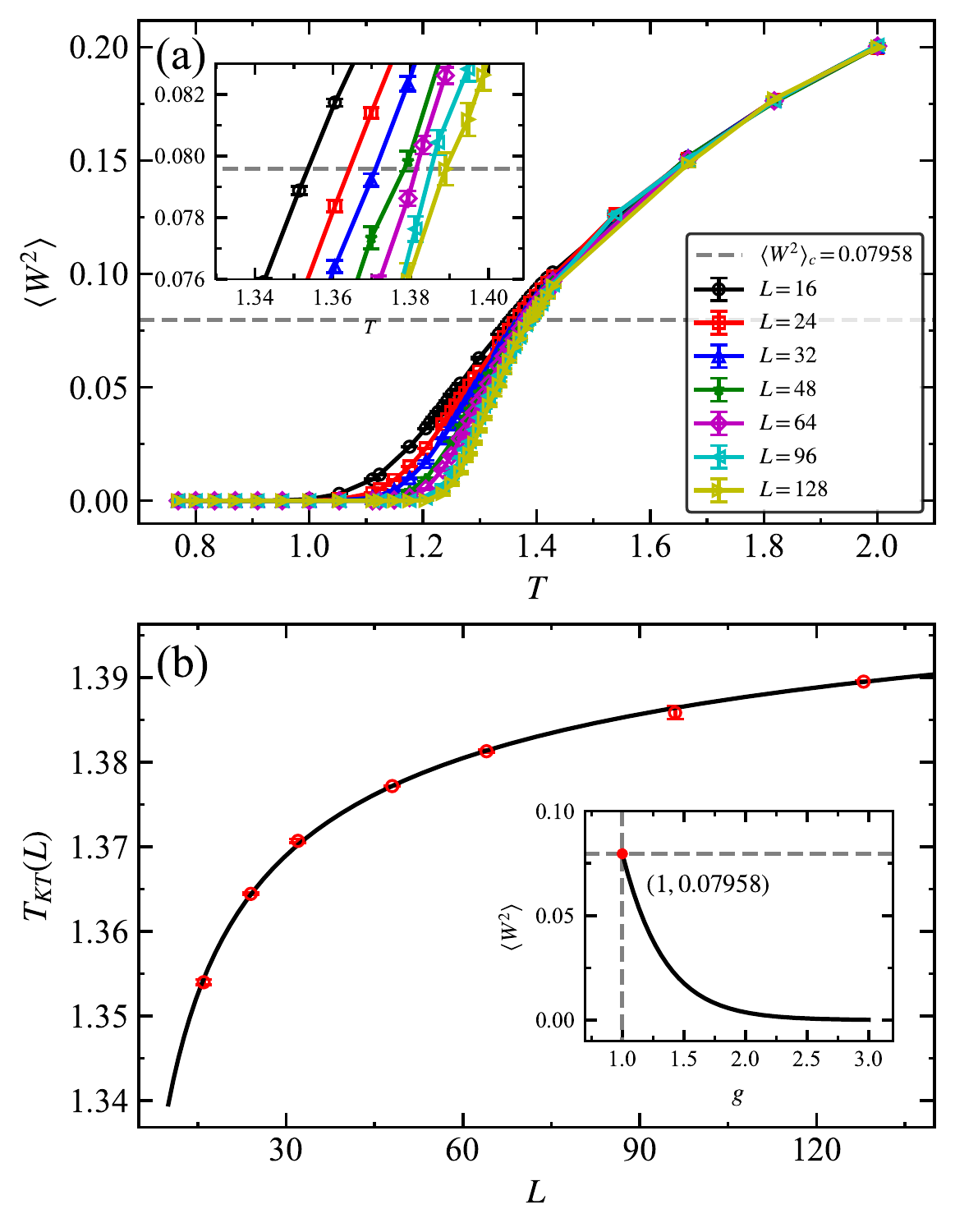}
\caption{(a) MC results for winding number fluctuations as a function of $T$. The gray dashed line shows the critical winding number fluctuations $\langle W^2 \rangle_c=0.07958$, which is obtained from Eq.~\eqref{eq:eq11} with $g_c=1$ at the transition point. Inset is a zoom-in for the $1.33\leqslant T\leqslant 1.41$ region. (b) Finite-size scaling for the estimated transition temperature as a function of system size. The finite-size $T_{KT}(L)$ data points are obtained from (a). The black curve shows the fit to Eq.~\eqref{eq:eq15}. The extrapolation to the thermodynamic limit gives $T_{KT}=1.425(1)$. The inset shows $\langle W^2 \rangle$ as a function of $g$ according to the relation in Eq.~\eqref{eq:eq11}.}
\label{fig:2}
\end{figure}

 \begin{figure}[htp!]
\centering
\includegraphics[width=1\columnwidth]{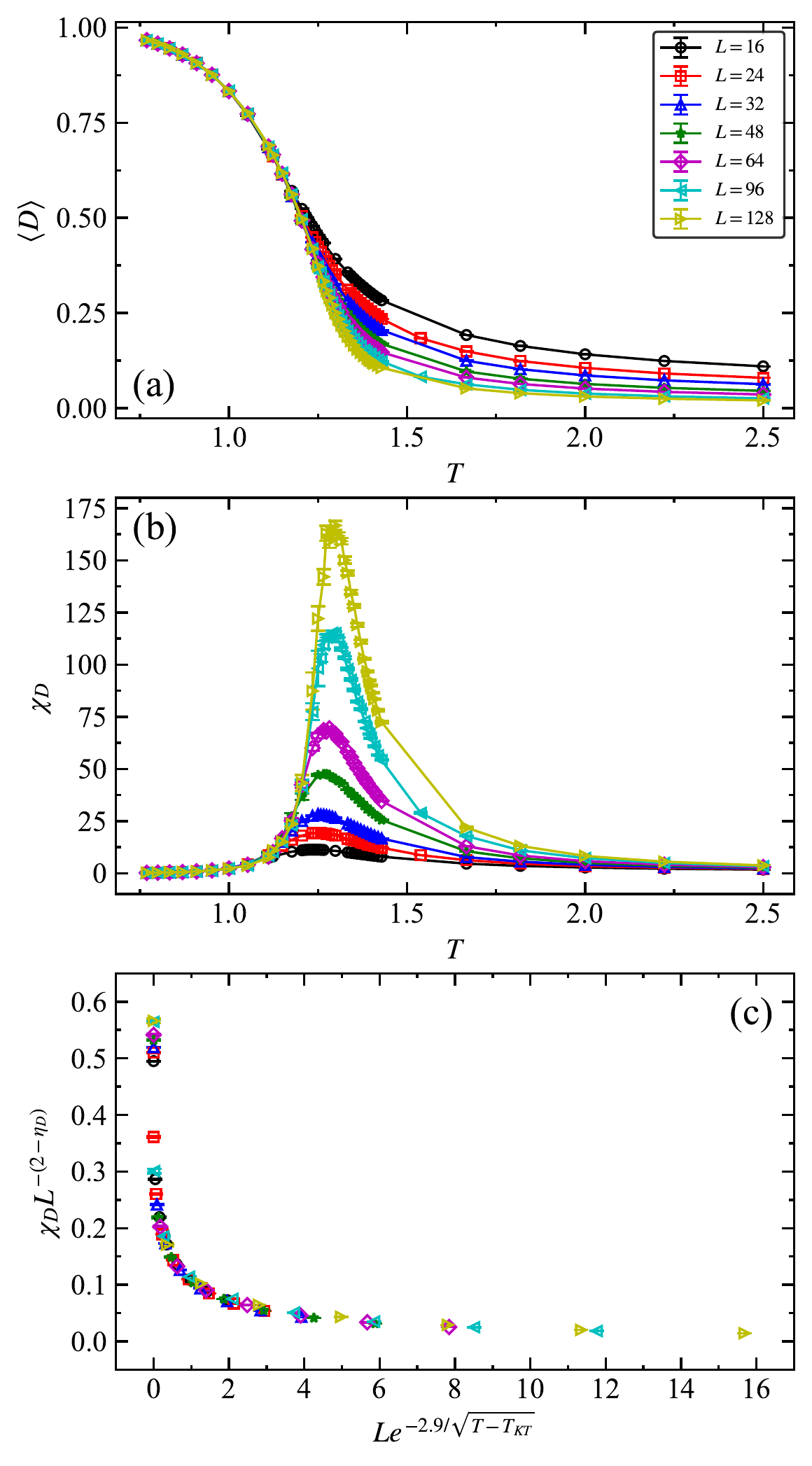}
\caption{(a) The nematic order parameter $\langle D\rangle$ as a function of temperature for different system sizes. (b) Susceptibility $\chi_D$ of the nematic order parameter defined in Eq.~\eqref{eq:eq4}. Data collapse is performed in (c) the critical phase with $T>T_{KT}$. Here we use $\eta_D=1$ and $T_{KT} = 1.425$, and all the data point nicely collapse onto a single curve.}
\label{fig:3}
\end{figure}

\section{MC simulation results}
\label{sec:v}
We present our MC simulation results in this section. It contains results for observables from which we can precisely estimate the critical temperature $T_{KT}$: the winding number fluctuations (Sec.~\ref{sec:va}) and the essential singularity of the nematic susceptibility in the KT transition (Sec.~\ref{sec:vb}). Section~\ref{sec:vc} presents results for different correlation functions in the high-temperature critical phase, confirming the field theoretical analysis presented in Sec.~\ref{sec:iv}.

\subsection{Winding number fluctuations}
\label{sec:va}
The numerical results for the winding number fluctuations $\langle W^2 \rangle$ as a function of temperature $T$ of the classical loop model are shown in Fig.~\ref{fig:2} (a). We simulate system sizes up to $L=128$ for this measurement. These data directly provide the temperature dependence (albeit on finite size) of the Coulomb gas constant, which will later be compared with other estimates of $g(T)$.
At the transition point, the analysis of Sec.~\ref{sec:iv} predicts the critical Coulomb gas constant $g_c$ to be $1$, corresponding to the critical winding number fluctuations $\langle W^2 \rangle_c=0.07958$  [from Eq.(~\ref{eq:eq11}), also see inset of Fig.~\ref{fig:2}(b)]. The predicted critical value $\langle W^2 \rangle_c$ is shown as the gray dashed horizontal line in Fig.~\ref{fig:2} (a) and in its inset.

We estimate the transition temperature $T_{KT}(L)$ for each system size as the temperature at which the winding number fluctuations cross the critical value, which is in turn estimated from a linear fit of the data points near $\langle W^{2}\rangle_c$ [Fig.~\ref{fig:2} (a)inset].
This estimate has an obvious finite-size dependence. To determine the transition temperature $T_{KT}$ in the thermodynamic limit, we use the following finite-size scaling relation for a KT transition~\cite{archambaultUniversal1998,atchisonFinite2019,wilkins2020}:
\begin{equation}
    \frac{1}{T_{KT}(L)} = \frac{1}{T_{KT}} + \frac{C}{\log(L/L_{0})^{2}},
    \label{eq:eq15}
\end{equation}
where $C$ is a constant.  By fitting the estimated $T_{KT}(L)$ in Fig.~\ref{fig:2} (b) with Eq.~\eqref{eq:eq15}, we obtain $T_{KT} = 1.425(1)$.

\begin{figure}[htp!]
	\centering
	\includegraphics[width=1\columnwidth]{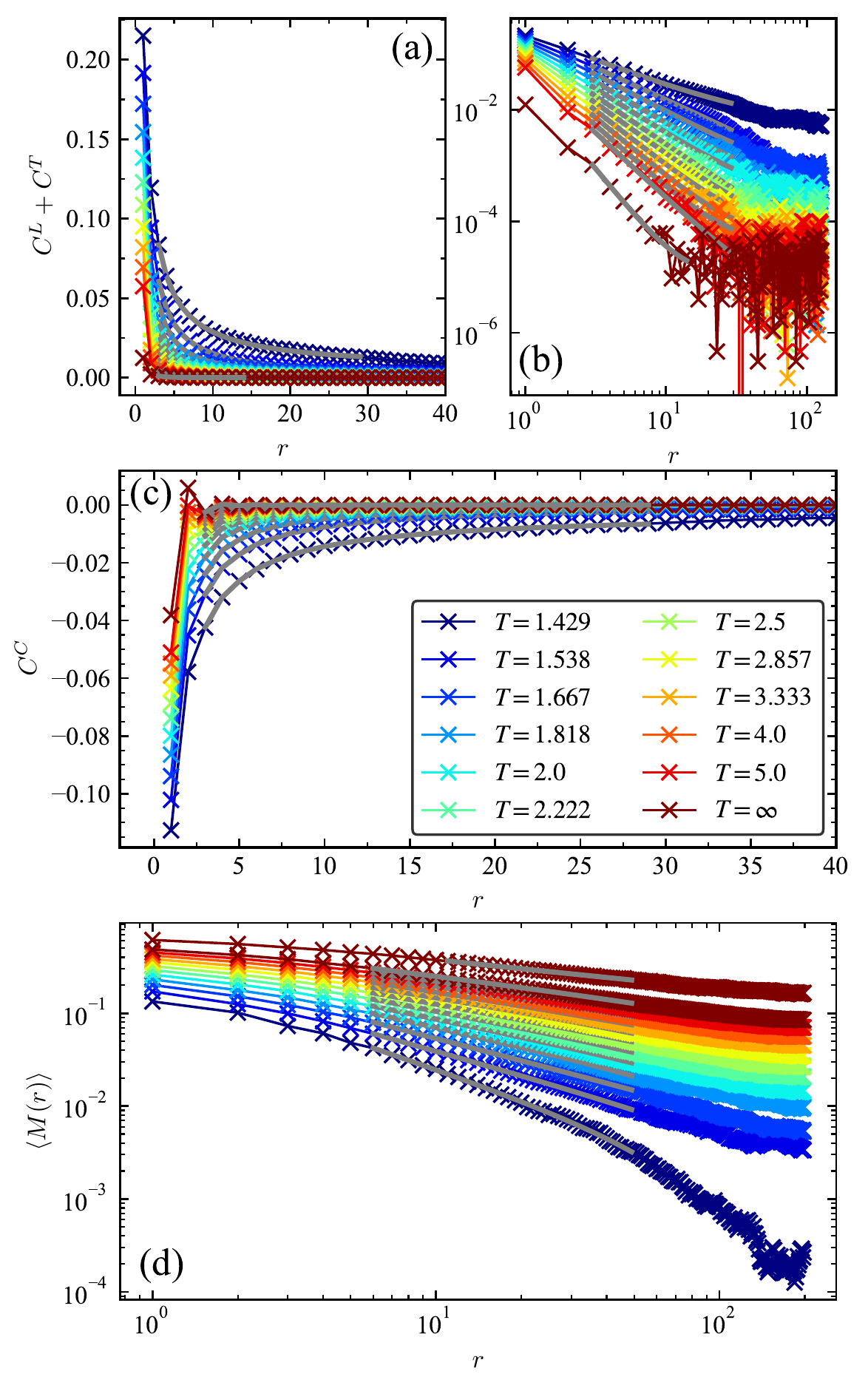}
	\caption{Equal-time loop-segment correlation functions (a) $C^L + C^T$ as a function of $r$, (b) the log-log plot for $|C^L + C^T|$ (absolute value is used to correct for very small negative values occurring at large $r$, large $T$ due to statistical fluctuations caused by the finite Monte Carlo sampling). (c) The crossed correlations $C^C$ (Eq.~\eqref{eq:eq7}), and (d) The log-log plot of the monomer correlations $\langle M(r) \rangle$ in Eq.~\eqref{eq:eq9}. The system size is $L=256$ for the loop-segment correlators and $L=400$ for monomer correlations. These data correspond to the high-temperature critical phase, that is temperatures above the estimated $T_{KT} = 1.425$. Gray curves are power-law fits (in their respective fitting range) according to the scaling form $B'/r^{1/g} + C$ for $C^L + C^T$, $(-1)^r A'/r^{\alpha_S} + B'/r^{\alpha_U} + C$ for $C^C$, and $B'/r^g+C$ for $M(r)$. In all cases, we add a constant to the power-law fits to account for a small non-vanishing value of correlators at large-distance in our finite-size Monte Carlo simulations.} 
	\label{fig:4}
\end{figure}

\begin{figure}[htp!]
	\centering
	\includegraphics[width=1\columnwidth]{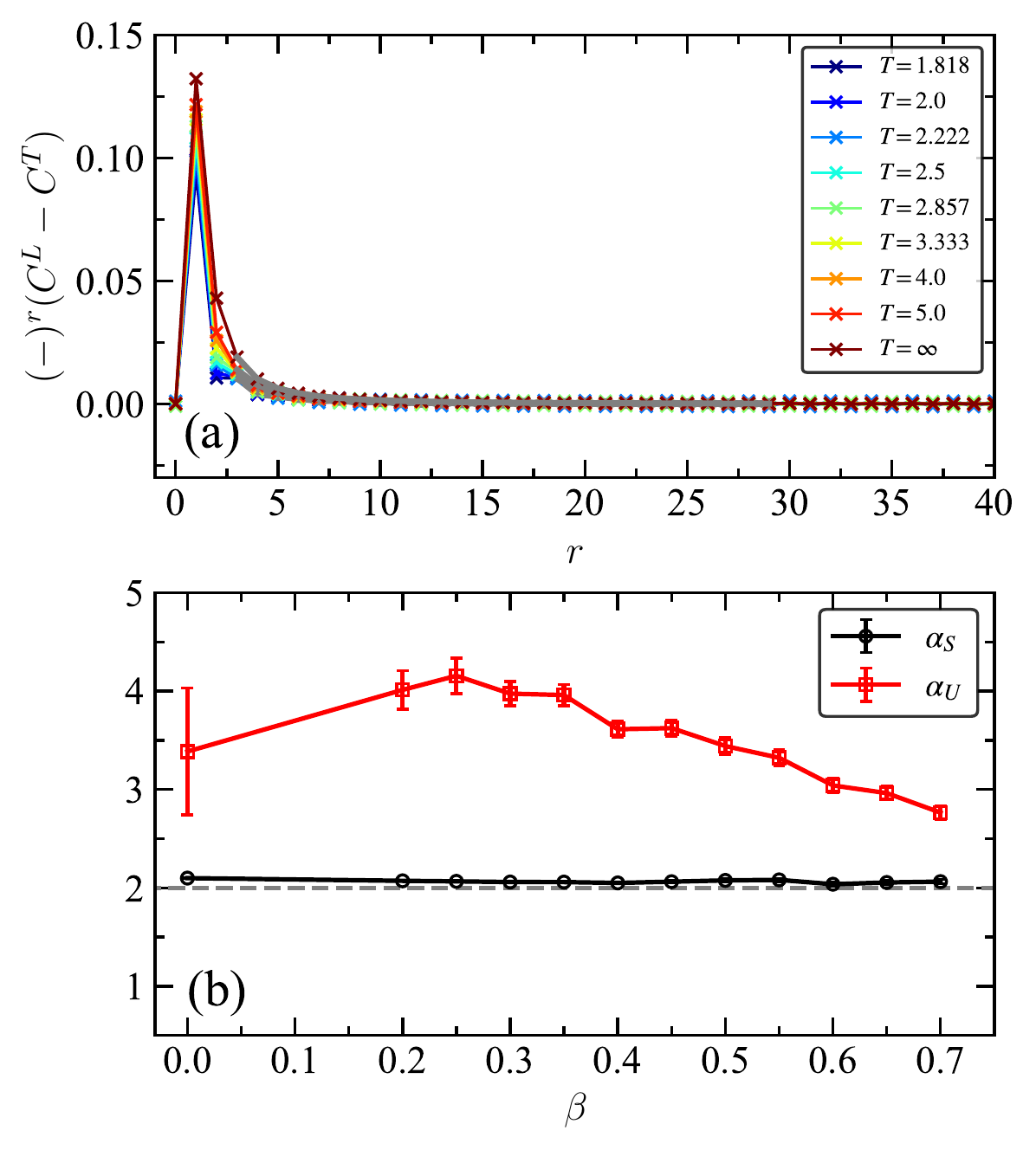}
	\caption{
 (a) The equal-time loop-segment correlation functions of $(-)^r(C^L - C^T)$ for $L=256$. A weak staggered part can be observed in this representation (particularly visible for the lowest temperatures at short distances), signaling a small uniform component for $(C^L - C^T)$. The gray curves fit $(C^L - C^T)$ to the form $(-1)^r A'/r^{\alpha_S} + B'/r^{\alpha_U} + C$. (b) The two exponents $\alpha_S$ and $\alpha_U$ obtained from this fit.}
	\label{fig:6}
\end{figure}

\begin{figure}[htp!]
	\centering
	\includegraphics[width=1\columnwidth]{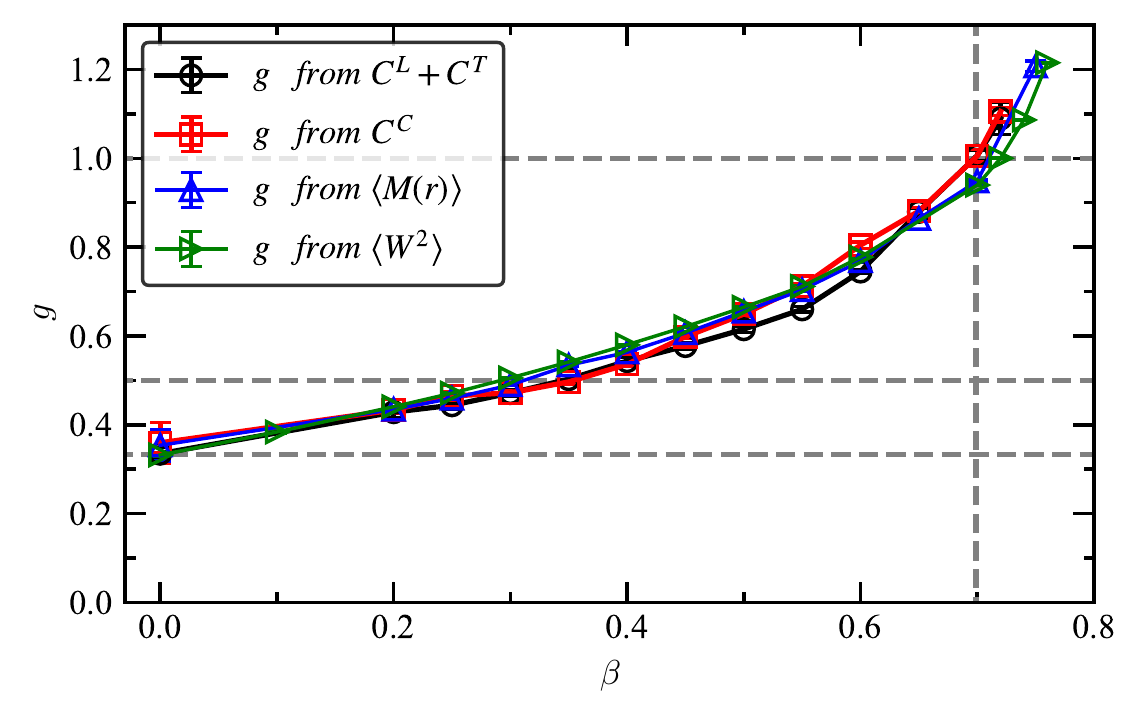}
	\caption{The Coulomb gas constant $g$ obtained from Fig.~\ref{fig:4} (b), (c), (d), and $\langle W^2 \rangle$. The vertical gray dash line indicates the transition point $\beta_{KT}=1/T_{KT} \simeq 0.7$. The upper gray dash line in the horizontal direction denotes the critical value $g_c=1$; the middle one at $g=1/2$ corresponds to the case where the vertex and dipolar terms have the same contribution to scaling, the lowest gray dash lines indicates the infinite temperature value $g=1/3$.} 
	\label{fig:5}
\end{figure}

\subsection{Nematic order parameter and susceptibility}
\label{sec:vb}

Our results for the finite-temperature behavior of the nematic order parameter $\langle D \rangle$ defined in Eq.(~\ref{eq:orderparameter})  are presented in Fig.~\ref{fig:3} (a) for different system sizes. We clearly observe the existence of a nematic phase at low temperature, where $\langle D \rangle$ takes a finite value, and a high-temperature phase where $\langle D \rangle$ vanishes relatively slowly as system sizes increase. The associated nematic susceptibility $\chi_D$ [Eq.~\eqref{eq:eq4}], represented in Fig.~\ref{fig:3} (b), shows a clear diverging peak (with system size) in the temperature range where $\langle D \rangle$ starts to vanish.

There is a clear shift in the temperature of the peak in $\chi_D$ as system size varies, as this can be used to perform a data collapse in order to cross-validate the transition temperature $T_{KT}$ obtained in the previous section. In the vicinity of the KT transition, we indeed expect that the susceptibility $\chi_D$ obeys the scaling behavior~\cite{Kosterlitz74} 
\begin{equation}
\chi_D \sim L^{2-\eta_D}f\left(L \exp(-\frac{K}{\sqrt{T-T_{KT}}})\right)
\end{equation}
for $T>T_{KT}$ where $K$ is a constant and $\eta_D=1/g_c=1$ the anomalous dimension~\cite{alet2005Interacting}. Such data collapse has been used in the literature to determine the Kosterlitz-Thouless transition temperature $T_{KT}$ in many 2D systems such as the 2D XY model~\cite{archambaultUniversal1998}, magnetic thin films~\cite{atchisonFinite2019}, triangular lattice transverse field Ising model~\cite{isakov2003interplay}, or for the pairing transition in various 2D fermionic lattice models~\cite{paiva2004critical,chenFermi2021,costaPhonon2018,jiangMonte2022}. We use the data in the $T>T_{KT}$ region to rescale the $y$ axis as $\chi_D L^{-(2-\eta_D)}$ and the $x$ axis as $L \exp(-\frac{K}{\sqrt{T-T_{KT}}})$ as shown in Fig.~\ref{fig:3} (c). We obtain that values $T_{KT}=1.425, \eta_D=1$ provide a good data collapse, resulting in a good agreement with the $T_{KT}$ obtained in Fig.~\ref{fig:2} (b).

\subsection{Correlation functions}
\label{sec:vc}

The height description of the loop-segment correlations given in Eqs.~\eqref{eq:eq12}, \eqref{eq:eq13} and \eqref{eq:eq14} suggests that the correlators have two contributions - from the vertex and the dipolar part ~\cite{moessner2004,lukyanov_long-distance_2003,falco2013,sandvik2006correlations}. 
We show the correlators calculated from the Monte Carlo methods for $L=256$ system and present the fits to their expected forms in Appendix~\ref{sec:appII}. Below we present a simpler approach to fitting - by considering  combinations of the correlators that separate out the vertex and dipolar terms.

We first consider the sum $C^L + C^T = \langle n_{\ohordimer}(0) n_{\ohordimer}({\bf r})\rangle+\langle n_{\overdimer}(0) n_{\overdimer}({\bf r})\rangle-\frac{1}{2}$ which should contain only a vertex contribution $2B/(x^2+y^2)^{1/2g}$. This combination for the direction ${\bf r}=(r,0)$ is shown in Figs.~\ref{fig:4}(a) and (b) in linear and log scales respectively. Consistent with the expected form, the combination shows a power-law scaling with distance $r$ with an exponent (slope in the log plot) that increases with temperature. The estimated value of $g$ from this combination is discussed further below.

We then consider the crossed correlators $C^C = \langle n_{\ohordimer}(0) n_{\overdimer}({\bf r})\rangle-1/2$ along the direction  ${\bf r}\equiv (r,0)$; where the correlation is expected to be dominated by the vertex term according to Eq. \ref{eq:eq14}.
Monte Carlo estimates of $C^{C}(r)$ for different temperatures are shown in Fig.~\ref{fig:4} (c).
The crossed correlators show expected power law scaling at large distances but with a possible oscillatory subleading correction that affects the short distance correlations, which is visible at higher temperatures. We tentatively attribute this effect to further subleading terms that do not cancel for $y=0$ and are not included in Eq.~\ref{eq:eq14}. This is confirmed by a fit to the form $(-1)^r A'/r^{\alpha_S} + B'/r^{\alpha_U} + C$ for $C^{C}(r)$ in Fig.~\ref{fig:4} (c), where we find that the amplitude $A'$ of the oscillating term is always small, and almost vanishing as the temperature lowers towards the critical point ({\it e.g.} $A'\sim 0.03$ for $T=4$ and $A'\sim 0.002$ for $T=1.3$). The results of this fit for $\alpha_U$ allow to estimate $g$ as obtained from Eq.~\ref{eq:eq14} and as presented below.

Next, we study the combination $C^L - C^T = \langle n_{\ohordimer}(0) n_{\ohordimer}({\bf r})\rangle-\langle n_{\overdimer}(0) n_{\overdimer}({\bf r})\rangle$ which, based on  Eqs.~\eqref{eq:eq12} and \eqref{eq:eq13}, is expected to have a purely staggered dipolar contribution $(-)^r/r^2$. 
Our results for $(-)^r(C^L - C^T)({\bf r})$ in the ${\bf r}=(r,0)$ direction are presented in Fig.~\ref{fig:6} (a). 
We observe that there is a small but non-vanishing uniform component in the numerical data (which appears as a staggered part in Fig.~\ref{fig:6} (a) due to the $(-)^r$ factor). To account for this, we fit $C^L-C^T$ to a form $(-1)^r A'/r^{\alpha_S} + B'/r^{\alpha_U} + C$. The constant $C$ accounts for a non-zero value of this correlator present only at temperatures close to the phase transition, which we attribute to the finite sizes used in our Monte Carlo simulations. Here $\alpha_U$ is meant to describe a subleading correction to the vertex part not included in Eqs.~\eqref{eq:eq12} and \eqref{eq:eq13}. The estimates for $\alpha_S,\alpha_U$ are presented in Fig.~\ref{fig:6} (b), where we find that $\alpha_S$ is very close to the predicted value $2$ all along the high-temperature critical phase, and $\alpha_U > \alpha_S$ confirming the subleading nature of this uniform correction. If we fit the data fixing $\alpha_U$ to be zero, the exponent $\alpha_S$ is always larger than its expected value of $2$.


Finally, the monomer-monomer correlator $M(r)$ in Eq.~\eqref{eq:eq9} should decay only with the vertex contribution $1/r^{g}$~\cite{hikihara1998correlation,alet2005Interacting,sandvik2006correlations,dabholkar2022reentrance}. We present the monomer correlations $M(r)$ at different temperatures in Fig.~\ref{fig:4} (d) for $L=400$ (within the directed loop algorithm, we can get good statistics for $M(r)$ for larger systems than for loop-segment correlators). The log-log plot shows a clear power-law decay above the Kosterlitz-Thouless transition temperature. 

 We now collect, in Fig.~\ref{fig:5}, the estimates of the Coulomb gas constant obtained from the fits to the correlators $C^{L}+C^{T}$ (from Fig.~\ref{fig:4}(a) and (b)), $C^{C}$ (from Fig.~\ref{fig:4}(c)) and $M(r)$ (from Fig.~\ref{fig:4}(d)) as well as from the winding number fluctuations $\langle W^2 \rangle$ in Fig.~\ref{fig:2}. 
 Fig.~\ref{fig:5} shows the temperature dependence of $g$ as a function of inverse temperature $\beta=1/T$. We find that as $\beta$ increases ({\it i.e.} as the temperature decreases), the Coulomb gas constant increases from its infinite temperature value $g=1/3$, which is consistent with the expectation that attractive interactions tend to {\it stiffen} the loops. 
 As the temperature decreases from the $T=\infty$ point to finite but high temperature, the dipolar part in Eqs.~\eqref{eq:eq12}, \eqref{eq:eq13} and \eqref{eq:eq14} dominates down to a temperature $T \approx 3$, below which $g>1/2$ and the vertex part takes over down to $T_{KT}\simeq 1.425$ $(\beta_{KT} \simeq 0.7)$ where $g=1$. 
 The various estimates of $g$ are overall in good agreement (we note the $g$ value obtained from $C^{C}$ is less accurate due to the subleading oscillations at high temperature) with each other, and consistent with the theoretical expectations of Sec.~\ref{sec:iv}.
 
\section{Discussion and conclusions}
\label{sec:vi}

In this work, we investigated the finite-temperature phase diagram of a classical model of fully-packed loops on the square lattice with attractive local interactions between loop segments. With the help of a directed-loop Monte Carlo algorithm and a field theoretical analysis based on a height description of loop configurations, we are able to locate the finite temperature Kosterlitz-Thouless transition, separating a critical phase at $T>T_{KT}$ and a nematic phase below $T_{KT}$. We find that in the loop model the anomalous dimension at the KT transition $\eta_D=1$, is four times larger than that in the classical dimer model~\cite{alet2005Interacting}. 
The high-temperature critical phase is fully characterized by the temperature dependence of the Coulomb gas constant presented in Fig.~\ref{fig:5}, which is obtained using several different concurrent estimates. 

An interesting, closely related system to consider would be a similar classical model, but with {\it repulsive} interactions ($V>0$) between fully packed loops, favoring large-winding sectors. Analogous repulsive interactions in the dimer model result in a continuous phase transition from a critical to staggered phase, which has been argued to be in the two-dimensional Ising universality class~\cite{wilkins2020}.

We connect our results to the quantum loop model on the square lattice. From our analysis, we expect that the QLM on the square lattice should also host a critical phase at any sufficiently high-temperature parametrized by a Coulomb gas constant $g(T/t,V/t)$ which depends on temperature and potential energy, similar to the quantum dimer model~\cite{dabholkar2022reentrance}. In general, a high-temperature critical phase can be found in the constrained entropic scaling regime of several strongly constrained quantum systems, which can extend down to low temperatures (see Ref.~\cite{castelnovoHigh2006} for an extensive discussion).  At large negative ratio of potential to kinetic energy ($V/t\ll0$), the QLM hosts a nematic ground-state. From our results, we  conclude that the finite-temperature phase transition to the nematic phase in the QLM should occur as a Kosterlitz-phase transition that can be described using the same analysis provided here.
The QLM also hosts a plaquette ground-state in a finite range of $-0.35 \lesssim V/t \ll 1$~\cite{ranFully2022}. We believe that the finite-temperature phase transition to this plaquette phase should be of KT type too, with an effective action described by Eq.~\eqref{eq:S} but with {\it negative} $v$, as the two plaquette ground-states have average height $\bar{h}=0,\frac{1}{2}$. It would be interesting to find a classical model with a similar phase transition and low-temperature phase. 
Finally, we note that the directed loop algorithm that we use can be directly implemented as a new move~\cite{dabholkar2022reentrance} within the sweeping cluster algorithm~\cite{yanSweeping2019,yanGlobal2022} for the QLM, allowing the study of its finite-temperature phase diagram fully taking into account the loop constraints and winding fluctuations. 

Rydberg atom arrays form a new type of platform where constraints (due to the Rydberg blockade) play an important role to determine the ground-state phase diagram, with a rich variety of phases observed \cite{Browaeys20,Celi20,Verresen21,schollQuantum2021,rhineQuantum2021,yanTriangular2022}. To the best of our knowledge, the finite-temperature phase transitions out of these phases has not been studied experimentally so far. It would be interesting to see where the finite-temperature critical phase that we find here could be relevant in some experimental regimes where the fully-packed constraint is a relevant approximation in Rydberg atom arrays.

{\it{Acknowledgments ---}} We acknowledge support from the ANR/RGC Joint Research Scheme sponsored by Research Grants Council of Hong Kong SAR of China (Project No. A\_HKU703/22) and French National Research Agency (grant ANR-22-CE30-0042-01). XXR, ZY and ZYM further acknowledge the support from the Research Grants Council of Hong Kong SAR of China (Project Nos. 17301420, 17301721, AoE/P-701/20, 17309822, HKU C7037-22G), and BD, GJS and FA the support from the joint PhD program between CNRS and IISER Pune, as well as the grant NanoX ANR-17-EURE-0009 in the framework of the French “Programme des Investissements d’Avenir”.
The research of JR is supported by the Huawei Young Talents Program at IHES.
We acknowledge the use of HPC resources from CALMIP (grants
2022-P0677 and 2023-P0677), GENCI (projects
A0110500225 and A0130500225), the HPC2021 system under the Information Technology Services, the Blackbody high-performance computing system at the Department of Physics, University of Hong Kong and Param Brahma computing facility at IISER Pune.”. GJS and BD thank K. Damle for useful discussions as well as TIFR, Mumbai for hospitality during the completion of this work.

\bibliographystyle{apsrev4-2}
\bibliography{main}

\begin{thebibliography}{108}%
\makeatletter
\providecommand \@ifxundefined [1]{%
 \@ifx{#1\undefined}
}%
\providecommand \@ifnum [1]{%
 \ifnum #1\expandafter \@firstoftwo
 \else \expandafter \@secondoftwo
 \fi
}%
\providecommand \@ifx [1]{%
 \ifx #1\expandafter \@firstoftwo
 \else \expandafter \@secondoftwo
 \fi
}%
\providecommand \natexlab [1]{#1}%
\providecommand \enquote  [1]{``#1''}%
\providecommand \bibnamefont  [1]{#1}%
\providecommand \bibfnamefont [1]{#1}%
\providecommand \citenamefont [1]{#1}%
\providecommand \href@noop [0]{\@secondoftwo}%
\providecommand \href [0]{\begingroup \@sanitize@url \@href}%
\providecommand \@href[1]{\@@startlink{#1}\@@href}%
\providecommand \@@href[1]{\endgroup#1\@@endlink}%
\providecommand \@sanitize@url [0]{\catcode `\\12\catcode `\$12\catcode
  `\&12\catcode `\#12\catcode `\^12\catcode `\_12\catcode `\%12\relax}%
\providecommand \@@startlink[1]{}%
\providecommand \@@endlink[0]{}%
\providecommand \url  [0]{\begingroup\@sanitize@url \@url }%
\providecommand \@url [1]{\endgroup\@href {#1}{\urlprefix }}%
\providecommand \urlprefix  [0]{URL }%
\providecommand \Eprint [0]{\href }%
\providecommand \doibase [0]{https://doi.org/}%
\providecommand \selectlanguage [0]{\@gobble}%
\providecommand \bibinfo  [0]{\@secondoftwo}%
\providecommand \bibfield  [0]{\@secondoftwo}%
\providecommand \translation [1]{[#1]}%
\providecommand \BibitemOpen [0]{}%
\providecommand \bibitemStop [0]{}%
\providecommand \bibitemNoStop [0]{.\EOS\space}%
\providecommand \EOS [0]{\spacefactor3000\relax}%
\providecommand \BibitemShut  [1]{\csname bibitem#1\endcsname}%
\let\auto@bib@innerbib\@empty
\bibitem [{\citenamefont {Moessner}\ and\ \citenamefont
  {Sondhi}(2001{\natexlab{a}})}]{moessnerResonating2001}%
  \BibitemOpen
  \bibfield  {author} {\bibinfo {author} {\bibfnamefont {R.}~\bibnamefont
  {Moessner}}\ and\ \bibinfo {author} {\bibfnamefont {S.~L.}\ \bibnamefont
  {Sondhi}},\ }\href {https://doi.org/10.1103/PhysRevLett.86.1881} {\bibfield
  {journal} {\bibinfo  {journal} {Phys. Rev. Lett.}\ }\textbf {\bibinfo
  {volume} {86}},\ \bibinfo {pages} {1881} (\bibinfo {year}
  {2001}{\natexlab{a}})}\BibitemShut {NoStop}%
\bibitem [{\citenamefont {Moessner}\ and\ \citenamefont
  {Sondhi}(2001{\natexlab{b}})}]{moessnerIsing2001}%
  \BibitemOpen
  \bibfield  {author} {\bibinfo {author} {\bibfnamefont {R.}~\bibnamefont
  {Moessner}}\ and\ \bibinfo {author} {\bibfnamefont {S.~L.}\ \bibnamefont
  {Sondhi}},\ }\href {https://doi.org/10.1103/PhysRevB.63.224401} {\bibfield
  {journal} {\bibinfo  {journal} {Phys. Rev. B}\ }\textbf {\bibinfo {volume}
  {63}},\ \bibinfo {pages} {224401} (\bibinfo {year}
  {2001}{\natexlab{b}})}\BibitemShut {NoStop}%
\bibitem [{\citenamefont {Bramwell}\ and\ \citenamefont
  {Gingras}(2001)}]{bramwell}%
  \BibitemOpen
  \bibfield  {author} {\bibinfo {author} {\bibfnamefont {S.~T.}\ \bibnamefont
  {Bramwell}}\ and\ \bibinfo {author} {\bibfnamefont {M.~J.~P.}\ \bibnamefont
  {Gingras}},\ }\href {https://doi.org/10.1126/science.1064761} {\bibfield
  {journal} {\bibinfo  {journal} {Science}\ }\textbf {\bibinfo {volume}
  {294}},\ \bibinfo {pages} {1495} (\bibinfo {year} {2001})},\ \Eprint
  {https://arxiv.org/abs/https://www.science.org/doi/pdf/10.1126/science.1064761}
  {https://www.science.org/doi/pdf/10.1126/science.1064761} \BibitemShut
  {NoStop}%
\bibitem [{\citenamefont {{Semeghini}}\ \emph {et~al.}(2021)\citenamefont
  {{Semeghini}}, \citenamefont {{Levine}}, \citenamefont {{Keesling}},
  \citenamefont {{Ebadi}}, \citenamefont {{Wang}}, \citenamefont {{Bluvstein}},
  \citenamefont {{Verresen}}, \citenamefont {{Pichler}}, \citenamefont
  {{Kalinowski}}, \citenamefont {{Samajdar}}, \citenamefont {{Omran}},
  \citenamefont {{Sachdev}}, \citenamefont {{Vishwanath}}, \citenamefont
  {{Greiner}}, \citenamefont {{Vuleti{\'c}}},\ and\ \citenamefont
  {{Lukin}}}]{semeghiniProbing2021}%
  \BibitemOpen
  \bibfield  {author} {\bibinfo {author} {\bibfnamefont {G.}~\bibnamefont
  {{Semeghini}}}, \bibinfo {author} {\bibfnamefont {H.}~\bibnamefont
  {{Levine}}}, \bibinfo {author} {\bibfnamefont {A.}~\bibnamefont
  {{Keesling}}}, \bibinfo {author} {\bibfnamefont {S.}~\bibnamefont {{Ebadi}}},
  \bibinfo {author} {\bibfnamefont {T.~T.}\ \bibnamefont {{Wang}}}, \bibinfo
  {author} {\bibfnamefont {D.}~\bibnamefont {{Bluvstein}}}, \bibinfo {author}
  {\bibfnamefont {R.}~\bibnamefont {{Verresen}}}, \bibinfo {author}
  {\bibfnamefont {H.}~\bibnamefont {{Pichler}}}, \bibinfo {author}
  {\bibfnamefont {M.}~\bibnamefont {{Kalinowski}}}, \bibinfo {author}
  {\bibfnamefont {R.}~\bibnamefont {{Samajdar}}}, \bibinfo {author}
  {\bibfnamefont {A.}~\bibnamefont {{Omran}}}, \bibinfo {author} {\bibfnamefont
  {S.}~\bibnamefont {{Sachdev}}}, \bibinfo {author} {\bibfnamefont
  {A.}~\bibnamefont {{Vishwanath}}}, \bibinfo {author} {\bibfnamefont
  {M.}~\bibnamefont {{Greiner}}}, \bibinfo {author} {\bibfnamefont
  {V.}~\bibnamefont {{Vuleti{\'c}}}},\ and\ \bibinfo {author} {\bibfnamefont
  {M.~D.}\ \bibnamefont {{Lukin}}},\ }\href
  {https://doi.org/10.1126/science.abi8794} {\bibfield  {journal} {\bibinfo
  {journal} {Science}\ }\textbf {\bibinfo {volume} {374}},\ \bibinfo {pages}
  {1242} (\bibinfo {year} {2021})}\BibitemShut {NoStop}%
\bibitem [{\citenamefont {Samajdar}\ \emph {et~al.}(2021)\citenamefont
  {Samajdar}, \citenamefont {Ho}, \citenamefont {Pichler}, \citenamefont
  {Lukin},\ and\ \citenamefont {Sachdev}}]{rhineQuantum2021}%
  \BibitemOpen
  \bibfield  {author} {\bibinfo {author} {\bibfnamefont {R.}~\bibnamefont
  {Samajdar}}, \bibinfo {author} {\bibfnamefont {W.~W.}\ \bibnamefont {Ho}},
  \bibinfo {author} {\bibfnamefont {H.}~\bibnamefont {Pichler}}, \bibinfo
  {author} {\bibfnamefont {M.~D.}\ \bibnamefont {Lukin}},\ and\ \bibinfo
  {author} {\bibfnamefont {S.}~\bibnamefont {Sachdev}},\ }\href
  {https://doi.org/10.1073/pnas.2015785118} {\bibfield  {journal} {\bibinfo
  {journal} {Proceedings of the National Academy of Sciences}\ }\textbf
  {\bibinfo {volume} {118}},\ \bibinfo {pages} {e2015785118} (\bibinfo {year}
  {2021})}\BibitemShut {NoStop}%
\bibitem [{\citenamefont {Ebadi}\ \emph {et~al.}(2021)\citenamefont {Ebadi},
  \citenamefont {Wang}, \citenamefont {Levine}, \citenamefont {Keesling},
  \citenamefont {Semeghini}, \citenamefont {Omran}, \citenamefont {Bluvstein},
  \citenamefont {Samajdar}, \citenamefont {Pichler}, \citenamefont {Ho},
  \citenamefont {Choi}, \citenamefont {Sachdev}, \citenamefont {Greiner},
  \citenamefont {Vuleti\'c},\ and\ \citenamefont {Lukin}}]{ebadiQuantum2021}%
  \BibitemOpen
  \bibfield  {author} {\bibinfo {author} {\bibfnamefont {S.}~\bibnamefont
  {Ebadi}}, \bibinfo {author} {\bibfnamefont {T.~T.}\ \bibnamefont {Wang}},
  \bibinfo {author} {\bibfnamefont {H.}~\bibnamefont {Levine}}, \bibinfo
  {author} {\bibfnamefont {A.}~\bibnamefont {Keesling}}, \bibinfo {author}
  {\bibfnamefont {G.}~\bibnamefont {Semeghini}}, \bibinfo {author}
  {\bibfnamefont {A.}~\bibnamefont {Omran}}, \bibinfo {author} {\bibfnamefont
  {D.}~\bibnamefont {Bluvstein}}, \bibinfo {author} {\bibfnamefont
  {R.}~\bibnamefont {Samajdar}}, \bibinfo {author} {\bibfnamefont
  {H.}~\bibnamefont {Pichler}}, \bibinfo {author} {\bibfnamefont {W.~W.}\
  \bibnamefont {Ho}}, \bibinfo {author} {\bibfnamefont {S.}~\bibnamefont
  {Choi}}, \bibinfo {author} {\bibfnamefont {S.}~\bibnamefont {Sachdev}},
  \bibinfo {author} {\bibfnamefont {M.}~\bibnamefont {Greiner}}, \bibinfo
  {author} {\bibfnamefont {V.}~\bibnamefont {Vuleti\'c}},\ and\ \bibinfo
  {author} {\bibfnamefont {M.~D.}\ \bibnamefont {Lukin}},\ }\href
  {https://doi.org/10.1038/s41586-021-03582-4} {\bibfield  {journal} {\bibinfo
  {journal} {Nature}\ }\textbf {\bibinfo {volume} {595}},\ \bibinfo {pages}
  {227} (\bibinfo {year} {2021})}\BibitemShut {NoStop}%
\bibitem [{\citenamefont {{Yan}}\ \emph {et~al.}(2022)\citenamefont {{Yan}},
  \citenamefont {{Samajdar}}, \citenamefont {{Wang}}, \citenamefont
  {{Sachdev}},\ and\ \citenamefont {{Meng}}}]{yanTriangular2022}%
  \BibitemOpen
  \bibfield  {author} {\bibinfo {author} {\bibfnamefont {Z.}~\bibnamefont
  {{Yan}}}, \bibinfo {author} {\bibfnamefont {R.}~\bibnamefont {{Samajdar}}},
  \bibinfo {author} {\bibfnamefont {Y.-C.}\ \bibnamefont {{Wang}}}, \bibinfo
  {author} {\bibfnamefont {S.}~\bibnamefont {{Sachdev}}},\ and\ \bibinfo
  {author} {\bibfnamefont {Z.~Y.}\ \bibnamefont {{Meng}}},\ }\href
  {https://doi.org/https://doi.org/10.1038/s41467-022-33431-5} {\bibfield
  {journal} {\bibinfo  {journal} {Nat. Commun.}\ }\textbf {\bibinfo {volume}
  {13}},\ \bibinfo {pages} {5799} (\bibinfo {year} {2022})}\BibitemShut
  {NoStop}%
\bibitem [{\citenamefont {Rokhsar}\ and\ \citenamefont
  {Kivelson}(1988)}]{rokhsarSuperconductivity1988}%
  \BibitemOpen
  \bibfield  {author} {\bibinfo {author} {\bibfnamefont {D.~S.}\ \bibnamefont
  {Rokhsar}}\ and\ \bibinfo {author} {\bibfnamefont {S.~A.}\ \bibnamefont
  {Kivelson}},\ }\href {https://doi.org/10.1103/PhysRevLett.61.2376} {\bibfield
   {journal} {\bibinfo  {journal} {Phys. Rev. Lett.}\ }\textbf {\bibinfo
  {volume} {61}},\ \bibinfo {pages} {2376} (\bibinfo {year}
  {1988})}\BibitemShut {NoStop}%
\bibitem [{\citenamefont {{Blunt}}\ \emph {et~al.}(2008)\citenamefont
  {{Blunt}}, \citenamefont {{Russell}}, \citenamefont
  {{Gim{\'e}nez-L{\'o}pez}}, \citenamefont {{Garrahan}}, \citenamefont {{Lin}},
  \citenamefont {{Schr{\"o}der}}, \citenamefont {{Champness}},\ and\
  \citenamefont {{Beton}}}]{blunt08}%
  \BibitemOpen
  \bibfield  {author} {\bibinfo {author} {\bibfnamefont {M.~O.}\ \bibnamefont
  {{Blunt}}}, \bibinfo {author} {\bibfnamefont {J.~C.}\ \bibnamefont
  {{Russell}}}, \bibinfo {author} {\bibfnamefont {M.~d.~C.}\ \bibnamefont
  {{Gim{\'e}nez-L{\'o}pez}}}, \bibinfo {author} {\bibfnamefont {J.~P.}\
  \bibnamefont {{Garrahan}}}, \bibinfo {author} {\bibfnamefont
  {X.}~\bibnamefont {{Lin}}}, \bibinfo {author} {\bibfnamefont
  {M.}~\bibnamefont {{Schr{\"o}der}}}, \bibinfo {author} {\bibfnamefont
  {N.~R.}\ \bibnamefont {{Champness}}},\ and\ \bibinfo {author} {\bibfnamefont
  {P.~H.}\ \bibnamefont {{Beton}}},\ }\href
  {https://doi.org/10.1126/science.1163338} {\bibfield  {journal} {\bibinfo
  {journal} {Science}\ }\textbf {\bibinfo {volume} {322}},\ \bibinfo {pages}
  {1077} (\bibinfo {year} {2008})}\BibitemShut {NoStop}%
\bibitem [{\citenamefont {Gruzberg}\ \emph {et~al.}(1999)\citenamefont
  {Gruzberg}, \citenamefont {Ludwig},\ and\ \citenamefont {Read}}]{gruzberg99}%
  \BibitemOpen
  \bibfield  {author} {\bibinfo {author} {\bibfnamefont {I.~A.}\ \bibnamefont
  {Gruzberg}}, \bibinfo {author} {\bibfnamefont {A.~W.~W.}\ \bibnamefont
  {Ludwig}},\ and\ \bibinfo {author} {\bibfnamefont {N.}~\bibnamefont {Read}},\
  }\href {https://doi.org/10.1103/PhysRevLett.82.4524} {\bibfield  {journal}
  {\bibinfo  {journal} {Phys. Rev. Lett.}\ }\textbf {\bibinfo {volume} {82}},\
  \bibinfo {pages} {4524} (\bibinfo {year} {1999})}\BibitemShut {NoStop}%
\bibitem [{\citenamefont {Read}\ and\ \citenamefont {Saleur}(2001)}]{read01}%
  \BibitemOpen
  \bibfield  {author} {\bibinfo {author} {\bibfnamefont {N.}~\bibnamefont
  {Read}}\ and\ \bibinfo {author} {\bibfnamefont {H.}~\bibnamefont {Saleur}},\
  }\href {https://doi.org/https://doi.org/10.1016/S0550-3213(01)00395-9}
  {\bibfield  {journal} {\bibinfo  {journal} {Nuclear Physics B}\ }\textbf
  {\bibinfo {volume} {613}},\ \bibinfo {pages} {409} (\bibinfo {year}
  {2001})}\BibitemShut {NoStop}%
\bibitem [{\citenamefont {Fendley}(2008)}]{fendley08}%
  \BibitemOpen
  \bibfield  {author} {\bibinfo {author} {\bibfnamefont {P.}~\bibnamefont
  {Fendley}},\ }\href
  {https://doi.org/https://doi.org/10.1016/j.aop.2008.04.011} {\bibfield
  {journal} {\bibinfo  {journal} {Annals of Physics}\ }\textbf {\bibinfo
  {volume} {323}},\ \bibinfo {pages} {3113} (\bibinfo {year}
  {2008})}\BibitemShut {NoStop}%
\bibitem [{\citenamefont {Nahum}\ \emph {et~al.}(2011)\citenamefont {Nahum},
  \citenamefont {Chalker}, \citenamefont {Serna}, \citenamefont {Ortu\~no},\
  and\ \citenamefont {Somoza}}]{Nahum1}%
  \BibitemOpen
  \bibfield  {author} {\bibinfo {author} {\bibfnamefont {A.}~\bibnamefont
  {Nahum}}, \bibinfo {author} {\bibfnamefont {J.~T.}\ \bibnamefont {Chalker}},
  \bibinfo {author} {\bibfnamefont {P.}~\bibnamefont {Serna}}, \bibinfo
  {author} {\bibfnamefont {M.}~\bibnamefont {Ortu\~no}},\ and\ \bibinfo
  {author} {\bibfnamefont {A.~M.}\ \bibnamefont {Somoza}},\ }\href
  {https://doi.org/10.1103/PhysRevLett.107.110601} {\bibfield  {journal}
  {\bibinfo  {journal} {Phys. Rev. Lett.}\ }\textbf {\bibinfo {volume} {107}},\
  \bibinfo {pages} {110601} (\bibinfo {year} {2011})}\BibitemShut {NoStop}%
\bibitem [{\citenamefont {Nahum}\ \emph {et~al.}(2015)\citenamefont {Nahum},
  \citenamefont {Chalker}, \citenamefont {Serna}, \citenamefont {Ortu\~no},\
  and\ \citenamefont {Somoza}}]{nahum15}%
  \BibitemOpen
  \bibfield  {author} {\bibinfo {author} {\bibfnamefont {A.}~\bibnamefont
  {Nahum}}, \bibinfo {author} {\bibfnamefont {J.~T.}\ \bibnamefont {Chalker}},
  \bibinfo {author} {\bibfnamefont {P.}~\bibnamefont {Serna}}, \bibinfo
  {author} {\bibfnamefont {M.}~\bibnamefont {Ortu\~no}},\ and\ \bibinfo
  {author} {\bibfnamefont {A.~M.}\ \bibnamefont {Somoza}},\ }\href
  {https://doi.org/10.1103/PhysRevX.5.041048} {\bibfield  {journal} {\bibinfo
  {journal} {Phys. Rev. X}\ }\textbf {\bibinfo {volume} {5}},\ \bibinfo {pages}
  {041048} (\bibinfo {year} {2015})}\BibitemShut {NoStop}%
\bibitem [{\citenamefont {Alet}\ \emph
  {et~al.}(2006{\natexlab{a}})\citenamefont {Alet}, \citenamefont {Misguich},
  \citenamefont {Pasquier}, \citenamefont {Moessner},\ and\ \citenamefont
  {Jacobsen}}]{AletUnconventional}%
  \BibitemOpen
  \bibfield  {author} {\bibinfo {author} {\bibfnamefont {F.}~\bibnamefont
  {Alet}}, \bibinfo {author} {\bibfnamefont {G.}~\bibnamefont {Misguich}},
  \bibinfo {author} {\bibfnamefont {V.}~\bibnamefont {Pasquier}}, \bibinfo
  {author} {\bibfnamefont {R.}~\bibnamefont {Moessner}},\ and\ \bibinfo
  {author} {\bibfnamefont {J.~L.}\ \bibnamefont {Jacobsen}},\ }\href
  {https://doi.org/10.1103/PhysRevLett.97.030403} {\bibfield  {journal}
  {\bibinfo  {journal} {Phys. Rev. Lett.}\ }\textbf {\bibinfo {volume} {97}},\
  \bibinfo {pages} {030403} (\bibinfo {year} {2006}{\natexlab{a}})}\BibitemShut
  {NoStop}%
\bibitem [{\citenamefont {Powell}\ and\ \citenamefont
  {Chalker}(2008)}]{PowellContinuumTheory}%
  \BibitemOpen
  \bibfield  {author} {\bibinfo {author} {\bibfnamefont {S.}~\bibnamefont
  {Powell}}\ and\ \bibinfo {author} {\bibfnamefont {J.~T.}\ \bibnamefont
  {Chalker}},\ }\href {https://doi.org/10.1103/PhysRevLett.101.155702}
  {\bibfield  {journal} {\bibinfo  {journal} {Phys. Rev. Lett.}\ }\textbf
  {\bibinfo {volume} {101}},\ \bibinfo {pages} {155702} (\bibinfo {year}
  {2008})}\BibitemShut {NoStop}%
\bibitem [{\citenamefont {Charrier}\ \emph {et~al.}(2008)\citenamefont
  {Charrier}, \citenamefont {Alet},\ and\ \citenamefont
  {Pujol}}]{AletGaugeTheory08}%
  \BibitemOpen
  \bibfield  {author} {\bibinfo {author} {\bibfnamefont {D.}~\bibnamefont
  {Charrier}}, \bibinfo {author} {\bibfnamefont {F.}~\bibnamefont {Alet}},\
  and\ \bibinfo {author} {\bibfnamefont {P.}~\bibnamefont {Pujol}},\ }\href
  {https://doi.org/10.1103/PhysRevLett.101.167205} {\bibfield  {journal}
  {\bibinfo  {journal} {Phys. Rev. Lett.}\ }\textbf {\bibinfo {volume} {101}},\
  \bibinfo {pages} {167205} (\bibinfo {year} {2008})}\BibitemShut {NoStop}%
\bibitem [{\citenamefont {Powell}\ and\ \citenamefont
  {Chalker}(2009)}]{PowellClassical2Quantum}%
  \BibitemOpen
  \bibfield  {author} {\bibinfo {author} {\bibfnamefont {S.}~\bibnamefont
  {Powell}}\ and\ \bibinfo {author} {\bibfnamefont {J.~T.}\ \bibnamefont
  {Chalker}},\ }\href {https://doi.org/10.1103/PhysRevB.80.134413} {\bibfield
  {journal} {\bibinfo  {journal} {Phys. Rev. B}\ }\textbf {\bibinfo {volume}
  {80}},\ \bibinfo {pages} {134413} (\bibinfo {year} {2009})}\BibitemShut
  {NoStop}%
\bibitem [{\citenamefont {Chen}\ \emph {et~al.}(2009)\citenamefont {Chen},
  \citenamefont {Gukelberger}, \citenamefont {Trebst}, \citenamefont {Alet},\
  and\ \citenamefont {Balents}}]{GangChen3DDimer}%
  \BibitemOpen
  \bibfield  {author} {\bibinfo {author} {\bibfnamefont {G.}~\bibnamefont
  {Chen}}, \bibinfo {author} {\bibfnamefont {J.}~\bibnamefont {Gukelberger}},
  \bibinfo {author} {\bibfnamefont {S.}~\bibnamefont {Trebst}}, \bibinfo
  {author} {\bibfnamefont {F.}~\bibnamefont {Alet}},\ and\ \bibinfo {author}
  {\bibfnamefont {L.}~\bibnamefont {Balents}},\ }\href
  {https://doi.org/10.1103/PhysRevB.80.045112} {\bibfield  {journal} {\bibinfo
  {journal} {Phys. Rev. B}\ }\textbf {\bibinfo {volume} {80}},\ \bibinfo
  {pages} {045112} (\bibinfo {year} {2009})}\BibitemShut {NoStop}%
\bibitem [{\citenamefont {Sreejith}\ and\ \citenamefont
  {Powell}(2015)}]{SreejithScalingDim}%
  \BibitemOpen
  \bibfield  {author} {\bibinfo {author} {\bibfnamefont {G.~J.}\ \bibnamefont
  {Sreejith}}\ and\ \bibinfo {author} {\bibfnamefont {S.}~\bibnamefont
  {Powell}},\ }\href {https://doi.org/10.1103/PhysRevB.92.184413} {\bibfield
  {journal} {\bibinfo  {journal} {Phys. Rev. B}\ }\textbf {\bibinfo {volume}
  {92}},\ \bibinfo {pages} {184413} (\bibinfo {year} {2015})}\BibitemShut
  {NoStop}%
\bibitem [{\citenamefont {Sreejith}\ \emph {et~al.}(2019)\citenamefont
  {Sreejith}, \citenamefont {Powell},\ and\ \citenamefont
  {Nahum}}]{SreejithSO5}%
  \BibitemOpen
  \bibfield  {author} {\bibinfo {author} {\bibfnamefont {G.~J.}\ \bibnamefont
  {Sreejith}}, \bibinfo {author} {\bibfnamefont {S.}~\bibnamefont {Powell}},\
  and\ \bibinfo {author} {\bibfnamefont {A.}~\bibnamefont {Nahum}},\ }\href
  {https://doi.org/10.1103/PhysRevLett.122.080601} {\bibfield  {journal}
  {\bibinfo  {journal} {Phys. Rev. Lett.}\ }\textbf {\bibinfo {volume} {122}},\
  \bibinfo {pages} {080601} (\bibinfo {year} {2019})}\BibitemShut {NoStop}%
\bibitem [{\citenamefont {Sreejith}\ and\ \citenamefont
  {Powell}(2014)}]{SreejithMonopoleFugacity}%
  \BibitemOpen
  \bibfield  {author} {\bibinfo {author} {\bibfnamefont {G.~J.}\ \bibnamefont
  {Sreejith}}\ and\ \bibinfo {author} {\bibfnamefont {S.}~\bibnamefont
  {Powell}},\ }\href {https://doi.org/10.1103/PhysRevB.89.014404} {\bibfield
  {journal} {\bibinfo  {journal} {Phys. Rev. B}\ }\textbf {\bibinfo {volume}
  {89}},\ \bibinfo {pages} {014404} (\bibinfo {year} {2014})}\BibitemShut
  {NoStop}%
\bibitem [{\citenamefont {Nienhuis}(2010)}]{nienhuis2010}%
  \BibitemOpen
  \bibfield  {author} {\bibinfo {author} {\bibfnamefont {B.}~\bibnamefont
  {Nienhuis}},\ }in\ \href@noop {} {\emph {\bibinfo {booktitle} {Exact methods
  in low-dimensional statistical physics and quantum computing}}},\ \bibinfo
  {editor} {edited by\ \bibinfo {editor} {\bibfnamefont {J.}~\bibnamefont
  {Jacobsen}}, \bibinfo {editor} {\bibfnamefont {S.}~\bibnamefont {Ouvry}},
  \bibinfo {editor} {\bibfnamefont {V.}~\bibnamefont {Pasquier}}, \bibinfo
  {editor} {\bibfnamefont {D.}~\bibnamefont {Serban}},\ and\ \bibinfo {editor}
  {\bibfnamefont {L.~F.}\ \bibnamefont {Cugliandolo}}}\ (\bibinfo  {publisher}
  {Oxford University Press},\ \bibinfo {address} {Oxford},\ \bibinfo {year}
  {2010})\ Chap.~\bibinfo {chapter} {6}, pp.\ \bibinfo {pages}
  {159--195}\BibitemShut {NoStop}%
\bibitem [{\citenamefont {Nienhuis}(1987)}]{nienhuis87}%
  \BibitemOpen
  \bibfield  {author} {\bibinfo {author} {\bibfnamefont {B.}~\bibnamefont
  {Nienhuis}},\ }in\ \href@noop {} {\emph {\bibinfo {booktitle} {Phase
  transitions and critical phenomena Vol. 11}}},\ \bibinfo {editor} {edited by\
  \bibinfo {editor} {\bibfnamefont {C.}~\bibnamefont {Domb}}\ and\ \bibinfo
  {editor} {\bibfnamefont {J.}~\bibnamefont {Lebowitz}}}\ (\bibinfo
  {publisher} {Academic Press},\ \bibinfo {year} {1987})\ Chap.~\bibinfo
  {chapter} {1}, pp.\ \bibinfo {pages} {1--53}\BibitemShut {NoStop}%
\bibitem [{\citenamefont {Jacobsen}(2009)}]{jacobsen09}%
  \BibitemOpen
  \bibfield  {author} {\bibinfo {author} {\bibfnamefont {J.~L.}\ \bibnamefont
  {Jacobsen}},\ }\bibinfo {title} {Conformal field theory applied to loop
  models},\ in\ \href {https://doi.org/10.1007/978-1-4020-9927-4_14} {\emph
  {\bibinfo {booktitle} {Polygons, Polyominoes and Polycubes}}},\ \bibinfo
  {editor} {edited by\ \bibinfo {editor} {\bibfnamefont {A.~J.}\ \bibnamefont
  {Guttman}}}\ (\bibinfo  {publisher} {Springer Netherlands},\ \bibinfo
  {address} {Dordrecht},\ \bibinfo {year} {2009})\ pp.\ \bibinfo {pages}
  {347--424}\BibitemShut {NoStop}%
\bibitem [{\citenamefont {Nahum}\ \emph {et~al.}(2013)\citenamefont {Nahum},
  \citenamefont {Serna}, \citenamefont {Somoza},\ and\ \citenamefont
  {Ortu\~no}}]{nahum13}%
  \BibitemOpen
  \bibfield  {author} {\bibinfo {author} {\bibfnamefont {A.}~\bibnamefont
  {Nahum}}, \bibinfo {author} {\bibfnamefont {P.}~\bibnamefont {Serna}},
  \bibinfo {author} {\bibfnamefont {A.~M.}\ \bibnamefont {Somoza}},\ and\
  \bibinfo {author} {\bibfnamefont {M.}~\bibnamefont {Ortu\~no}},\ }\href
  {https://doi.org/10.1103/PhysRevB.87.184204} {\bibfield  {journal} {\bibinfo
  {journal} {Phys. Rev. B}\ }\textbf {\bibinfo {volume} {87}},\ \bibinfo
  {pages} {184204} (\bibinfo {year} {2013})}\BibitemShut {NoStop}%
\bibitem [{\citenamefont {Kondev}\ \emph {et~al.}(1996)\citenamefont {Kondev},
  \citenamefont {de~Gier},\ and\ \citenamefont {Nienhuis}}]{Kondev1996}%
  \BibitemOpen
  \bibfield  {author} {\bibinfo {author} {\bibfnamefont {J.}~\bibnamefont
  {Kondev}}, \bibinfo {author} {\bibfnamefont {J.}~\bibnamefont {de~Gier}},\
  and\ \bibinfo {author} {\bibfnamefont {B.}~\bibnamefont {Nienhuis}},\ }\href
  {https://doi.org/10.1088/0305-4470/29/20/007} {\bibfield  {journal} {\bibinfo
   {journal} {Journal of Physics A: Mathematical and General}\ }\textbf
  {\bibinfo {volume} {29}},\ \bibinfo {pages} {6489} (\bibinfo {year}
  {1996})}\BibitemShut {NoStop}%
\bibitem [{\citenamefont {Temperley}\ and\ \citenamefont
  {Lieb}(1971)}]{temperley71}%
  \BibitemOpen
  \bibfield  {author} {\bibinfo {author} {\bibfnamefont {H.~N.~V.}\
  \bibnamefont {Temperley}}\ and\ \bibinfo {author} {\bibfnamefont {E.~H.}\
  \bibnamefont {Lieb}},\ }\href {https://doi.org/10.1098/rspa.1971.0067}
  {\bibfield  {journal} {\bibinfo  {journal} {Proc. Roy. Soc. Lond. A}\
  }\textbf {\bibinfo {volume} {322}},\ \bibinfo {pages} {251} (\bibinfo {year}
  {1971})}\BibitemShut {NoStop}%
\bibitem [{\citenamefont {{de Gennes}}(1972)}]{de_gennes72}%
  \BibitemOpen
  \bibfield  {author} {\bibinfo {author} {\bibfnamefont {P.}~\bibnamefont {{de
  Gennes}}},\ }\href
  {https://doi.org/https://doi.org/10.1016/0375-9601(72)90149-1} {\bibfield
  {journal} {\bibinfo  {journal} {Physics Letters A}\ }\textbf {\bibinfo
  {volume} {38}},\ \bibinfo {pages} {339} (\bibinfo {year} {1972})}\BibitemShut
  {NoStop}%
\bibitem [{\citenamefont {Jacobsen}\ \emph {et~al.}(2003)\citenamefont
  {Jacobsen}, \citenamefont {Read},\ and\ \citenamefont {Saleur}}]{jacobsen03}%
  \BibitemOpen
  \bibfield  {author} {\bibinfo {author} {\bibfnamefont {J.~L.}\ \bibnamefont
  {Jacobsen}}, \bibinfo {author} {\bibfnamefont {N.}~\bibnamefont {Read}},\
  and\ \bibinfo {author} {\bibfnamefont {H.}~\bibnamefont {Saleur}},\ }\href
  {https://doi.org/10.1103/PhysRevLett.90.090601} {\bibfield  {journal}
  {\bibinfo  {journal} {Phys. Rev. Lett.}\ }\textbf {\bibinfo {volume} {90}},\
  \bibinfo {pages} {090601} (\bibinfo {year} {2003})}\BibitemShut {NoStop}%
\bibitem [{\citenamefont {Cardy}(2005)}]{cardy05}%
  \BibitemOpen
  \bibfield  {author} {\bibinfo {author} {\bibfnamefont {J.}~\bibnamefont
  {Cardy}},\ }\href {https://doi.org/https://doi.org/10.1016/j.aop.2005.04.001}
  {\bibfield  {journal} {\bibinfo  {journal} {Annals of Physics}\ }\textbf
  {\bibinfo {volume} {318}},\ \bibinfo {pages} {81} (\bibinfo {year} {2005})},\
  \bibinfo {note} {special Issue}\BibitemShut {NoStop}%
\bibitem [{\citenamefont {Jacobsen}\ and\ \citenamefont
  {Alet}(2009)}]{jacobsen09b}%
  \BibitemOpen
  \bibfield  {author} {\bibinfo {author} {\bibfnamefont {J.~L.}\ \bibnamefont
  {Jacobsen}}\ and\ \bibinfo {author} {\bibfnamefont {F.}~\bibnamefont
  {Alet}},\ }\href {https://doi.org/10.1103/PhysRevLett.102.145702} {\bibfield
  {journal} {\bibinfo  {journal} {Phys. Rev. Lett.}\ }\textbf {\bibinfo
  {volume} {102}},\ \bibinfo {pages} {145702} (\bibinfo {year}
  {2009})}\BibitemShut {NoStop}%
\bibitem [{\citenamefont {Schwandt}\ \emph {et~al.}(2010)\citenamefont
  {Schwandt}, \citenamefont {Mambrini},\ and\ \citenamefont
  {Poilblanc}}]{schwandt2010}%
  \BibitemOpen
  \bibfield  {author} {\bibinfo {author} {\bibfnamefont {D.}~\bibnamefont
  {Schwandt}}, \bibinfo {author} {\bibfnamefont {M.}~\bibnamefont {Mambrini}},\
  and\ \bibinfo {author} {\bibfnamefont {D.}~\bibnamefont {Poilblanc}},\ }\href
  {https://doi.org/10.1103/PhysRevB.81.214413} {\bibfield  {journal} {\bibinfo
  {journal} {Phys. Rev. B}\ }\textbf {\bibinfo {volume} {81}},\ \bibinfo
  {pages} {214413} (\bibinfo {year} {2010})}\BibitemShut {NoStop}%
\bibitem [{\citenamefont {Barkema}\ and\ \citenamefont
  {Newman}(1998)}]{barkema1998monte}%
  \BibitemOpen
  \bibfield  {author} {\bibinfo {author} {\bibfnamefont {G.}~\bibnamefont
  {Barkema}}\ and\ \bibinfo {author} {\bibfnamefont {M.}~\bibnamefont
  {Newman}},\ }\href
  {https://journals.aps.org/pre/abstract/10.1103/PhysRevE.57.1155} {\bibfield
  {journal} {\bibinfo  {journal} {Physical Review E}\ }\textbf {\bibinfo
  {volume} {57}},\ \bibinfo {pages} {1155} (\bibinfo {year}
  {1998})}\BibitemShut {NoStop}%
\bibitem [{\citenamefont {Alet}\ \emph
  {et~al.}(2006{\natexlab{b}})\citenamefont {Alet}, \citenamefont {Ikhlef},
  \citenamefont {Jacobsen}, \citenamefont {Misguich},\ and\ \citenamefont
  {Pasquier}}]{alet2006classical}%
  \BibitemOpen
  \bibfield  {author} {\bibinfo {author} {\bibfnamefont {F.}~\bibnamefont
  {Alet}}, \bibinfo {author} {\bibfnamefont {Y.}~\bibnamefont {Ikhlef}},
  \bibinfo {author} {\bibfnamefont {J.~L.}\ \bibnamefont {Jacobsen}}, \bibinfo
  {author} {\bibfnamefont {G.}~\bibnamefont {Misguich}},\ and\ \bibinfo
  {author} {\bibfnamefont {V.}~\bibnamefont {Pasquier}},\ }\href
  {https://doi.org/10.1103/PhysRevE.74.041124} {\bibfield  {journal} {\bibinfo
  {journal} {Phys. Rev. E}\ }\textbf {\bibinfo {volume} {74}},\ \bibinfo
  {pages} {041124} (\bibinfo {year} {2006}{\natexlab{b}})}\BibitemShut
  {NoStop}%
\bibitem [{\citenamefont {Sandvik}\ and\ \citenamefont
  {Moessner}(2006)}]{sandvik2006correlations}%
  \BibitemOpen
  \bibfield  {author} {\bibinfo {author} {\bibfnamefont {A.~W.}\ \bibnamefont
  {Sandvik}}\ and\ \bibinfo {author} {\bibfnamefont {R.}~\bibnamefont
  {Moessner}},\ }\href
  {https://journals.aps.org/prb/abstract/10.1103/PhysRevB.73.144504} {\bibfield
   {journal} {\bibinfo  {journal} {Physical Review B}\ }\textbf {\bibinfo
  {volume} {73}},\ \bibinfo {pages} {144504} (\bibinfo {year}
  {2006})}\BibitemShut {NoStop}%
\bibitem [{\citenamefont {Sylju\aa{}sen}\ and\ \citenamefont
  {Sandvik}(2002)}]{syljuasenquantum2002}%
  \BibitemOpen
  \bibfield  {author} {\bibinfo {author} {\bibfnamefont {O.~F.}\ \bibnamefont
  {Sylju\aa{}sen}}\ and\ \bibinfo {author} {\bibfnamefont {A.~W.}\ \bibnamefont
  {Sandvik}},\ }\href {https://doi.org/10.1103/PhysRevE.66.046701} {\bibfield
  {journal} {\bibinfo  {journal} {Phys. Rev. E}\ }\textbf {\bibinfo {volume}
  {66}},\ \bibinfo {pages} {046701} (\bibinfo {year} {2002})}\BibitemShut
  {NoStop}%
\bibitem [{\citenamefont {Sylju{\aa}sen}\ and\ \citenamefont
  {Zvonarev}(2004)}]{syljuaasen2004directed}%
  \BibitemOpen
  \bibfield  {author} {\bibinfo {author} {\bibfnamefont {O.~F.}\ \bibnamefont
  {Sylju{\aa}sen}}\ and\ \bibinfo {author} {\bibfnamefont {M.}~\bibnamefont
  {Zvonarev}},\ }\href
  {https://journals.aps.org/pre/abstract/10.1103/PhysRevE.70.016118} {\bibfield
   {journal} {\bibinfo  {journal} {Physical Review E}\ }\textbf {\bibinfo
  {volume} {70}},\ \bibinfo {pages} {016118} (\bibinfo {year}
  {2004})}\BibitemShut {NoStop}%
\bibitem [{\citenamefont {Alet}\ and\ \citenamefont
  {S\o{}rensen}(2003)}]{aletdirected2003}%
  \BibitemOpen
  \bibfield  {author} {\bibinfo {author} {\bibfnamefont {F.}~\bibnamefont
  {Alet}}\ and\ \bibinfo {author} {\bibfnamefont {E.~S.}\ \bibnamefont
  {S\o{}rensen}},\ }\href {https://doi.org/10.1103/PhysRevE.68.026702}
  {\bibfield  {journal} {\bibinfo  {journal} {Phys. Rev. E}\ }\textbf {\bibinfo
  {volume} {68}},\ \bibinfo {pages} {026702} (\bibinfo {year}
  {2003})}\BibitemShut {NoStop}%
\bibitem [{\citenamefont {Kondev}\ and\ \citenamefont
  {Henley}(1996)}]{kondev96b}%
  \BibitemOpen
  \bibfield  {author} {\bibinfo {author} {\bibfnamefont {J.}~\bibnamefont
  {Kondev}}\ and\ \bibinfo {author} {\bibfnamefont {C.~L.}\ \bibnamefont
  {Henley}},\ }\href
  {https://doi.org/https://doi.org/10.1016/0550-3213(96)00064-8} {\bibfield
  {journal} {\bibinfo  {journal} {Nuclear Physics B}\ }\textbf {\bibinfo
  {volume} {464}},\ \bibinfo {pages} {540} (\bibinfo {year}
  {1996})}\BibitemShut {NoStop}%
\bibitem [{\citenamefont {{Moessner}}\ \emph {et~al.}(2004)\citenamefont
  {{Moessner}}, \citenamefont {{Tchernyshyov}},\ and\ \citenamefont
  {{Sondhi}}}]{moessner2004}%
  \BibitemOpen
  \bibfield  {author} {\bibinfo {author} {\bibfnamefont {R.}~\bibnamefont
  {{Moessner}}}, \bibinfo {author} {\bibfnamefont {O.}~\bibnamefont
  {{Tchernyshyov}}},\ and\ \bibinfo {author} {\bibfnamefont {S.~L.}\
  \bibnamefont {{Sondhi}}},\ }\href
  {https://doi.org/10.1023/B:JOSS.0000037247.54022.62} {\bibfield  {journal}
  {\bibinfo  {journal} {Journal of Statistical Physics}\ }\textbf {\bibinfo
  {volume} {116}},\ \bibinfo {pages} {755} (\bibinfo {year}
  {2004})}\BibitemShut {NoStop}%
\bibitem [{\citenamefont {Alet}\ \emph {et~al.}(2005)\citenamefont {Alet},
  \citenamefont {Jacobsen}, \citenamefont {Misguich}, \citenamefont {Pasquier},
  \citenamefont {Mila},\ and\ \citenamefont {Troyer}}]{alet2005Interacting}%
  \BibitemOpen
  \bibfield  {author} {\bibinfo {author} {\bibfnamefont {F.}~\bibnamefont
  {Alet}}, \bibinfo {author} {\bibfnamefont {J.~L.}\ \bibnamefont {Jacobsen}},
  \bibinfo {author} {\bibfnamefont {G.}~\bibnamefont {Misguich}}, \bibinfo
  {author} {\bibfnamefont {V.}~\bibnamefont {Pasquier}}, \bibinfo {author}
  {\bibfnamefont {F.}~\bibnamefont {Mila}},\ and\ \bibinfo {author}
  {\bibfnamefont {M.}~\bibnamefont {Troyer}},\ }\href
  {https://doi.org/10.1103/PhysRevLett.94.235702} {\bibfield  {journal}
  {\bibinfo  {journal} {Phys. Rev. Lett.}\ }\textbf {\bibinfo {volume} {94}},\
  \bibinfo {pages} {235702} (\bibinfo {year} {2005})}\BibitemShut {NoStop}%
\bibitem [{\citenamefont {Papanikolaou}\ \emph {et~al.}(2007)\citenamefont
  {Papanikolaou}, \citenamefont {Luijten},\ and\ \citenamefont
  {Fradkin}}]{papanikolaou2007quantum}%
  \BibitemOpen
  \bibfield  {author} {\bibinfo {author} {\bibfnamefont {S.}~\bibnamefont
  {Papanikolaou}}, \bibinfo {author} {\bibfnamefont {E.}~\bibnamefont
  {Luijten}},\ and\ \bibinfo {author} {\bibfnamefont {E.}~\bibnamefont
  {Fradkin}},\ }\href
  {https://journals.aps.org/prb/abstract/10.1103/PhysRevB.76.134514} {\bibfield
   {journal} {\bibinfo  {journal} {Physical Review B}\ }\textbf {\bibinfo
  {volume} {76}},\ \bibinfo {pages} {134514} (\bibinfo {year}
  {2007})}\BibitemShut {NoStop}%
\bibitem [{\citenamefont {Kundu}\ and\ \citenamefont
  {Damle}(2023)}]{kundu2023flux}%
  \BibitemOpen
  \bibfield  {author} {\bibinfo {author} {\bibfnamefont {S.}~\bibnamefont
  {Kundu}}\ and\ \bibinfo {author} {\bibfnamefont {K.}~\bibnamefont {Damle}},\
  }\href {https://arxiv.org/abs/2305.07012} {\bibinfo {title} {Flux
  fractionalization transition in two-dimensional dimer-loop models}} (\bibinfo
  {year} {2023}),\ \Eprint {https://arxiv.org/abs/2305.07012} {arXiv:2305.07012
  [cond-mat.stat-mech]} \BibitemShut {NoStop}%
\bibitem [{\citenamefont {Castelnovo}\ \emph {et~al.}(2005)\citenamefont
  {Castelnovo}, \citenamefont {Chamon}, \citenamefont {Mudry},\ and\
  \citenamefont {Pujol}}]{castelnovo2005}%
  \BibitemOpen
  \bibfield  {author} {\bibinfo {author} {\bibfnamefont {C.}~\bibnamefont
  {Castelnovo}}, \bibinfo {author} {\bibfnamefont {C.}~\bibnamefont {Chamon}},
  \bibinfo {author} {\bibfnamefont {C.}~\bibnamefont {Mudry}},\ and\ \bibinfo
  {author} {\bibfnamefont {P.}~\bibnamefont {Pujol}},\ }\href
  {https://doi.org/https://doi.org/10.1016/j.aop.2005.01.006} {\bibfield
  {journal} {\bibinfo  {journal} {Annals of Physics}\ }\textbf {\bibinfo
  {volume} {318}},\ \bibinfo {pages} {316} (\bibinfo {year}
  {2005})}\BibitemShut {NoStop}%
\bibitem [{\citenamefont {Balasubramanian}\ \emph {et~al.}(2022)\citenamefont
  {Balasubramanian}, \citenamefont {Galitski},\ and\ \citenamefont
  {Vishwanath}}]{Balasubramanian22}%
  \BibitemOpen
  \bibfield  {author} {\bibinfo {author} {\bibfnamefont {S.}~\bibnamefont
  {Balasubramanian}}, \bibinfo {author} {\bibfnamefont {V.}~\bibnamefont
  {Galitski}},\ and\ \bibinfo {author} {\bibfnamefont {A.}~\bibnamefont
  {Vishwanath}},\ }\href {https://doi.org/10.1103/PhysRevB.106.195127}
  {\bibfield  {journal} {\bibinfo  {journal} {Phys. Rev. B}\ }\textbf {\bibinfo
  {volume} {106}},\ \bibinfo {pages} {195127} (\bibinfo {year}
  {2022})}\BibitemShut {NoStop}%
\bibitem [{\citenamefont {Castelnovo}\ \emph {et~al.}(2007)\citenamefont
  {Castelnovo}, \citenamefont {Chamon}, \citenamefont {Mudry},\ and\
  \citenamefont {Pujol}}]{castelnovo2006}%
  \BibitemOpen
  \bibfield  {author} {\bibinfo {author} {\bibfnamefont {C.}~\bibnamefont
  {Castelnovo}}, \bibinfo {author} {\bibfnamefont {C.}~\bibnamefont {Chamon}},
  \bibinfo {author} {\bibfnamefont {C.}~\bibnamefont {Mudry}},\ and\ \bibinfo
  {author} {\bibfnamefont {P.}~\bibnamefont {Pujol}},\ }\href
  {https://doi.org/https://doi.org/10.1016/j.aop.2006.04.017} {\bibfield
  {journal} {\bibinfo  {journal} {Annals of Physics}\ }\textbf {\bibinfo
  {volume} {322}},\ \bibinfo {pages} {903} (\bibinfo {year}
  {2007})}\BibitemShut {NoStop}%
\bibitem [{\citenamefont {Henry}\ and\ \citenamefont
  {Roscilde}(2014)}]{henry14}%
  \BibitemOpen
  \bibfield  {author} {\bibinfo {author} {\bibfnamefont {L.-P.}\ \bibnamefont
  {Henry}}\ and\ \bibinfo {author} {\bibfnamefont {T.}~\bibnamefont
  {Roscilde}},\ }\href {https://doi.org/10.1103/PhysRevLett.113.027204}
  {\bibfield  {journal} {\bibinfo  {journal} {Phys. Rev. Lett.}\ }\textbf
  {\bibinfo {volume} {113}},\ \bibinfo {pages} {027204} (\bibinfo {year}
  {2014})}\BibitemShut {NoStop}%
\bibitem [{\citenamefont {Shannon}\ \emph {et~al.}(2004)\citenamefont
  {Shannon}, \citenamefont {Misguich},\ and\ \citenamefont
  {Penc}}]{shannonCyclic2004}%
  \BibitemOpen
  \bibfield  {author} {\bibinfo {author} {\bibfnamefont {N.}~\bibnamefont
  {Shannon}}, \bibinfo {author} {\bibfnamefont {G.}~\bibnamefont {Misguich}},\
  and\ \bibinfo {author} {\bibfnamefont {K.}~\bibnamefont {Penc}},\ }\href
  {https://doi.org/10.1103/PhysRevB.69.220403} {\bibfield  {journal} {\bibinfo
  {journal} {Phys. Rev. B}\ }\textbf {\bibinfo {volume} {69}},\ \bibinfo
  {pages} {220403} (\bibinfo {year} {2004})}\BibitemShut {NoStop}%
\bibitem [{\citenamefont {Sylju\aa{}sen}\ and\ \citenamefont
  {Chakravarty}(2006)}]{syljuasen06}%
  \BibitemOpen
  \bibfield  {author} {\bibinfo {author} {\bibfnamefont {O.~F.}\ \bibnamefont
  {Sylju\aa{}sen}}\ and\ \bibinfo {author} {\bibfnamefont {S.}~\bibnamefont
  {Chakravarty}},\ }\href {https://doi.org/10.1103/PhysRevLett.96.147004}
  {\bibfield  {journal} {\bibinfo  {journal} {Phys. Rev. Lett.}\ }\textbf
  {\bibinfo {volume} {96}},\ \bibinfo {pages} {147004} (\bibinfo {year}
  {2006})}\BibitemShut {NoStop}%
\bibitem [{\citenamefont {Plat}\ \emph {et~al.}(2015)\citenamefont {Plat},
  \citenamefont {Alet}, \citenamefont {Capponi},\ and\ \citenamefont
  {Totsuka}}]{Plat2015Magnetization}%
  \BibitemOpen
  \bibfield  {author} {\bibinfo {author} {\bibfnamefont {X.}~\bibnamefont
  {Plat}}, \bibinfo {author} {\bibfnamefont {F.}~\bibnamefont {Alet}}, \bibinfo
  {author} {\bibfnamefont {S.}~\bibnamefont {Capponi}},\ and\ \bibinfo {author}
  {\bibfnamefont {K.}~\bibnamefont {Totsuka}},\ }\href
  {https://doi.org/10.1103/PhysRevB.92.174402} {\bibfield  {journal} {\bibinfo
  {journal} {Phys. Rev. B}\ }\textbf {\bibinfo {volume} {92}},\ \bibinfo
  {pages} {174402} (\bibinfo {year} {2015})}\BibitemShut {NoStop}%
\bibitem [{\citenamefont {Roychowdhury}\ \emph {et~al.}(2015)\citenamefont
  {Roychowdhury}, \citenamefont {Bhattacharjee},\ and\ \citenamefont
  {Pollmann}}]{Roychowdhury2015topological}%
  \BibitemOpen
  \bibfield  {author} {\bibinfo {author} {\bibfnamefont {K.}~\bibnamefont
  {Roychowdhury}}, \bibinfo {author} {\bibfnamefont {S.}~\bibnamefont
  {Bhattacharjee}},\ and\ \bibinfo {author} {\bibfnamefont {F.}~\bibnamefont
  {Pollmann}},\ }\href {https://doi.org/10.1103/PhysRevB.92.075141} {\bibfield
  {journal} {\bibinfo  {journal} {Phys. Rev. B}\ }\textbf {\bibinfo {volume}
  {92}},\ \bibinfo {pages} {075141} (\bibinfo {year} {2015})}\BibitemShut
  {NoStop}%
\bibitem [{\citenamefont {Ran}\ \emph {et~al.}(2023)\citenamefont {Ran},
  \citenamefont {Yan}, \citenamefont {Wang}, \citenamefont {Rong},
  \citenamefont {Qi},\ and\ \citenamefont {Meng}}]{ranFully2022}%
  \BibitemOpen
  \bibfield  {author} {\bibinfo {author} {\bibfnamefont {X.}~\bibnamefont
  {Ran}}, \bibinfo {author} {\bibfnamefont {Z.}~\bibnamefont {Yan}}, \bibinfo
  {author} {\bibfnamefont {Y.-C.}\ \bibnamefont {Wang}}, \bibinfo {author}
  {\bibfnamefont {J.}~\bibnamefont {Rong}}, \bibinfo {author} {\bibfnamefont
  {Y.}~\bibnamefont {Qi}},\ and\ \bibinfo {author} {\bibfnamefont {Z.~Y.}\
  \bibnamefont {Meng}},\ }\href {https://doi.org/10.1103/PhysRevB.107.125134}
  {\bibfield  {journal} {\bibinfo  {journal} {Phys. Rev. B}\ }\textbf {\bibinfo
  {volume} {107}},\ \bibinfo {pages} {125134} (\bibinfo {year}
  {2023})}\BibitemShut {NoStop}%
\bibitem [{\citenamefont {Yan}\ \emph {et~al.}()\citenamefont {Yan},
  \citenamefont {Ran}, \citenamefont {Wang}, \citenamefont {Samajdar},
  \citenamefont {Rong}, \citenamefont {Sachdev}, \citenamefont {Qi},\ and\
  \citenamefont {Meng}}]{yanFully2022}%
  \BibitemOpen
  \bibfield  {author} {\bibinfo {author} {\bibfnamefont {Z.}~\bibnamefont
  {Yan}}, \bibinfo {author} {\bibfnamefont {X.}~\bibnamefont {Ran}}, \bibinfo
  {author} {\bibfnamefont {Y.-C.}\ \bibnamefont {Wang}}, \bibinfo {author}
  {\bibfnamefont {R.}~\bibnamefont {Samajdar}}, \bibinfo {author}
  {\bibfnamefont {J.}~\bibnamefont {Rong}}, \bibinfo {author} {\bibfnamefont
  {S.}~\bibnamefont {Sachdev}}, \bibinfo {author} {\bibfnamefont
  {Y.}~\bibnamefont {Qi}},\ and\ \bibinfo {author} {\bibfnamefont {Z.~Y.}\
  \bibnamefont {Meng}},\ }\href
  {https://ui.adsabs.harvard.edu/abs/2022arXiv220504472Y} {\ }\Eprint
  {https://arxiv.org/abs/2205.04472 (2022)} {arXiv:2205.04472 (2022)
  [cond-mat.str-el]} \BibitemShut {NoStop}%
\bibitem [{\citenamefont {Dabholkar}\ \emph {et~al.}(2022)\citenamefont
  {Dabholkar}, \citenamefont {Sreejith},\ and\ \citenamefont
  {Alet}}]{dabholkar2022reentrance}%
  \BibitemOpen
  \bibfield  {author} {\bibinfo {author} {\bibfnamefont {B.}~\bibnamefont
  {Dabholkar}}, \bibinfo {author} {\bibfnamefont {G.~J.}\ \bibnamefont
  {Sreejith}},\ and\ \bibinfo {author} {\bibfnamefont {F.}~\bibnamefont
  {Alet}},\ }\href {https://doi.org/10.1103/PhysRevB.106.205121} {\bibfield
  {journal} {\bibinfo  {journal} {Phys. Rev. B}\ }\textbf {\bibinfo {volume}
  {106}},\ \bibinfo {pages} {205121} (\bibinfo {year} {2022})}\BibitemShut
  {NoStop}%
\bibitem [{\citenamefont {Yan}\ \emph {et~al.}(2021{\natexlab{a}})\citenamefont
  {Yan}, \citenamefont {Zhou}, \citenamefont {Sylju\aa{}sen}, \citenamefont
  {Zhang}, \citenamefont {Yuan}, \citenamefont {Lou},\ and\ \citenamefont
  {Chen}}]{yanWidely2021}%
  \BibitemOpen
  \bibfield  {author} {\bibinfo {author} {\bibfnamefont {Z.}~\bibnamefont
  {Yan}}, \bibinfo {author} {\bibfnamefont {Z.}~\bibnamefont {Zhou}}, \bibinfo
  {author} {\bibfnamefont {O.~F.}\ \bibnamefont {Sylju\aa{}sen}}, \bibinfo
  {author} {\bibfnamefont {J.}~\bibnamefont {Zhang}}, \bibinfo {author}
  {\bibfnamefont {T.}~\bibnamefont {Yuan}}, \bibinfo {author} {\bibfnamefont
  {J.}~\bibnamefont {Lou}},\ and\ \bibinfo {author} {\bibfnamefont
  {Y.}~\bibnamefont {Chen}},\ }\href
  {https://doi.org/10.1103/PhysRevB.103.094421} {\bibfield  {journal} {\bibinfo
   {journal} {Phys. Rev. B}\ }\textbf {\bibinfo {volume} {103}},\ \bibinfo
  {pages} {094421} (\bibinfo {year} {2021}{\natexlab{a}})}\BibitemShut
  {NoStop}%
\bibitem [{\citenamefont {Yan}\ \emph {et~al.}(2022)\citenamefont {Yan},
  \citenamefont {Meng}, \citenamefont {Huse},\ and\ \citenamefont
  {Chan}}]{yanHeight2022}%
  \BibitemOpen
  \bibfield  {author} {\bibinfo {author} {\bibfnamefont {Z.}~\bibnamefont
  {Yan}}, \bibinfo {author} {\bibfnamefont {Z.~Y.}\ \bibnamefont {Meng}},
  \bibinfo {author} {\bibfnamefont {D.~A.}\ \bibnamefont {Huse}},\ and\
  \bibinfo {author} {\bibfnamefont {A.}~\bibnamefont {Chan}},\ }\href
  {https://doi.org/10.1103/PhysRevB.106.L041115} {\bibfield  {journal}
  {\bibinfo  {journal} {Phys. Rev. B}\ }\textbf {\bibinfo {volume} {106}},\
  \bibinfo {pages} {L041115} (\bibinfo {year} {2022})}\BibitemShut {NoStop}%
\bibitem [{\citenamefont {Verresen}\ and\ \citenamefont
  {Vishwanath}(2022)}]{Verresen22}%
  \BibitemOpen
  \bibfield  {author} {\bibinfo {author} {\bibfnamefont {R.}~\bibnamefont
  {Verresen}}\ and\ \bibinfo {author} {\bibfnamefont {A.}~\bibnamefont
  {Vishwanath}},\ }\href {https://doi.org/10.1103/PhysRevX.12.041029}
  {\bibfield  {journal} {\bibinfo  {journal} {Phys. Rev. X}\ }\textbf {\bibinfo
  {volume} {12}},\ \bibinfo {pages} {041029} (\bibinfo {year}
  {2022})}\BibitemShut {NoStop}%
\bibitem [{\citenamefont {Yan}\ \emph {et~al.}(2021{\natexlab{b}})\citenamefont
  {Yan}, \citenamefont {Wang}, \citenamefont {Ma}, \citenamefont {Qi},\ and\
  \citenamefont {Meng}}]{yanTopological2020}%
  \BibitemOpen
  \bibfield  {author} {\bibinfo {author} {\bibfnamefont {Z.}~\bibnamefont
  {Yan}}, \bibinfo {author} {\bibfnamefont {Y.-C.}\ \bibnamefont {Wang}},
  \bibinfo {author} {\bibfnamefont {N.}~\bibnamefont {Ma}}, \bibinfo {author}
  {\bibfnamefont {Y.}~\bibnamefont {Qi}},\ and\ \bibinfo {author}
  {\bibfnamefont {Z.~Y.}\ \bibnamefont {Meng}},\ }\href
  {https://doi.org/10.1038/s41535-021-00338-1} {\bibfield  {journal} {\bibinfo
  {journal} {npj Quantum Mater.}\ }\textbf {\bibinfo {volume} {6}},\ \bibinfo
  {pages} {39} (\bibinfo {year} {2021}{\natexlab{b}})}\BibitemShut {NoStop}%
\bibitem [{\citenamefont {Pollmann}\ \emph {et~al.}(2011)\citenamefont
  {Pollmann}, \citenamefont {Betouras}, \citenamefont {Shtengel},\ and\
  \citenamefont {Fulde}}]{Pollmann2011}%
  \BibitemOpen
  \bibfield  {author} {\bibinfo {author} {\bibfnamefont {F.}~\bibnamefont
  {Pollmann}}, \bibinfo {author} {\bibfnamefont {J.~J.}\ \bibnamefont
  {Betouras}}, \bibinfo {author} {\bibfnamefont {K.}~\bibnamefont {Shtengel}},\
  and\ \bibinfo {author} {\bibfnamefont {P.}~\bibnamefont {Fulde}},\ }\href
  {https://doi.org/10.1103/PhysRevB.83.155117} {\bibfield  {journal} {\bibinfo
  {journal} {Phys. Rev. B}\ }\textbf {\bibinfo {volume} {83}},\ \bibinfo
  {pages} {155117} (\bibinfo {year} {2011})}\BibitemShut {NoStop}%
\bibitem [{\citenamefont {Banerjee}\ \emph {et~al.}(2013)\citenamefont
  {Banerjee}, \citenamefont {Jiang}, \citenamefont {Widmer},\ and\
  \citenamefont {Wiese}}]{Banerjee2013}%
  \BibitemOpen
  \bibfield  {author} {\bibinfo {author} {\bibfnamefont {D.}~\bibnamefont
  {Banerjee}}, \bibinfo {author} {\bibfnamefont {F.-J.}\ \bibnamefont {Jiang}},
  \bibinfo {author} {\bibfnamefont {P.}~\bibnamefont {Widmer}},\ and\ \bibinfo
  {author} {\bibfnamefont {U.-J.}\ \bibnamefont {Wiese}},\ }\href
  {https://doi.org/10.1088/1742-5468/2013/12/P12010} {\bibfield  {journal}
  {\bibinfo  {journal} {Journal of Statistical Mechanics: Theory and
  Experiment}\ }\textbf {\bibinfo {volume} {2013}},\ \bibinfo {pages} {P12010}
  (\bibinfo {year} {2013})}\BibitemShut {NoStop}%
\bibitem [{\citenamefont {{Browaeys}}\ and\ \citenamefont
  {{Lahaye}}(2020)}]{Browaeys20}%
  \BibitemOpen
  \bibfield  {author} {\bibinfo {author} {\bibfnamefont {A.}~\bibnamefont
  {{Browaeys}}}\ and\ \bibinfo {author} {\bibfnamefont {T.}~\bibnamefont
  {{Lahaye}}},\ }\href {https://doi.org/10.1038/s41567-019-0733-z} {\bibfield
  {journal} {\bibinfo  {journal} {Nature Physics}\ }\textbf {\bibinfo {volume}
  {16}},\ \bibinfo {pages} {132} (\bibinfo {year} {2020})}\BibitemShut
  {NoStop}%
\bibitem [{\citenamefont {Glaetzle}\ \emph {et~al.}(2014)\citenamefont
  {Glaetzle}, \citenamefont {Dalmonte}, \citenamefont {Nath}, \citenamefont
  {Rousochatzakis}, \citenamefont {Moessner},\ and\ \citenamefont
  {Zoller}}]{Glaetzle14}%
  \BibitemOpen
  \bibfield  {author} {\bibinfo {author} {\bibfnamefont {A.~W.}\ \bibnamefont
  {Glaetzle}}, \bibinfo {author} {\bibfnamefont {M.}~\bibnamefont {Dalmonte}},
  \bibinfo {author} {\bibfnamefont {R.}~\bibnamefont {Nath}}, \bibinfo {author}
  {\bibfnamefont {I.}~\bibnamefont {Rousochatzakis}}, \bibinfo {author}
  {\bibfnamefont {R.}~\bibnamefont {Moessner}},\ and\ \bibinfo {author}
  {\bibfnamefont {P.}~\bibnamefont {Zoller}},\ }\href
  {https://doi.org/10.1103/PhysRevX.4.041037} {\bibfield  {journal} {\bibinfo
  {journal} {Phys. Rev. X}\ }\textbf {\bibinfo {volume} {4}},\ \bibinfo {pages}
  {041037} (\bibinfo {year} {2014})}\BibitemShut {NoStop}%
\bibitem [{\citenamefont {Celi}\ \emph {et~al.}(2020)\citenamefont {Celi},
  \citenamefont {Vermersch}, \citenamefont {Viyuela}, \citenamefont {Pichler},
  \citenamefont {Lukin},\ and\ \citenamefont {Zoller}}]{Celi20}%
  \BibitemOpen
  \bibfield  {author} {\bibinfo {author} {\bibfnamefont {A.}~\bibnamefont
  {Celi}}, \bibinfo {author} {\bibfnamefont {B.}~\bibnamefont {Vermersch}},
  \bibinfo {author} {\bibfnamefont {O.}~\bibnamefont {Viyuela}}, \bibinfo
  {author} {\bibfnamefont {H.}~\bibnamefont {Pichler}}, \bibinfo {author}
  {\bibfnamefont {M.~D.}\ \bibnamefont {Lukin}},\ and\ \bibinfo {author}
  {\bibfnamefont {P.}~\bibnamefont {Zoller}},\ }\href
  {https://doi.org/10.1103/PhysRevX.10.021057} {\bibfield  {journal} {\bibinfo
  {journal} {Phys. Rev. X}\ }\textbf {\bibinfo {volume} {10}},\ \bibinfo
  {pages} {021057} (\bibinfo {year} {2020})}\BibitemShut {NoStop}%
\bibitem [{\citenamefont {Verresen}\ \emph {et~al.}(2021)\citenamefont
  {Verresen}, \citenamefont {Lukin},\ and\ \citenamefont
  {Vishwanath}}]{Verresen21}%
  \BibitemOpen
  \bibfield  {author} {\bibinfo {author} {\bibfnamefont {R.}~\bibnamefont
  {Verresen}}, \bibinfo {author} {\bibfnamefont {M.~D.}\ \bibnamefont
  {Lukin}},\ and\ \bibinfo {author} {\bibfnamefont {A.}~\bibnamefont
  {Vishwanath}},\ }\href {https://doi.org/10.1103/PhysRevX.11.031005}
  {\bibfield  {journal} {\bibinfo  {journal} {Phys. Rev. X}\ }\textbf {\bibinfo
  {volume} {11}},\ \bibinfo {pages} {031005} (\bibinfo {year}
  {2021})}\BibitemShut {NoStop}%
\bibitem [{\citenamefont {Yan}\ \emph {et~al.}(2023)\citenamefont {Yan},
  \citenamefont {Wang}, \citenamefont {Samajdar}, \citenamefont {Sachdev},\
  and\ \citenamefont {Meng}}]{yanEmergent2023}%
  \BibitemOpen
  \bibfield  {author} {\bibinfo {author} {\bibfnamefont {Z.}~\bibnamefont
  {Yan}}, \bibinfo {author} {\bibfnamefont {Y.-C.}\ \bibnamefont {Wang}},
  \bibinfo {author} {\bibfnamefont {R.}~\bibnamefont {Samajdar}}, \bibinfo
  {author} {\bibfnamefont {S.}~\bibnamefont {Sachdev}},\ and\ \bibinfo {author}
  {\bibfnamefont {Z.~Y.}\ \bibnamefont {Meng}},\ }\href
  {https://doi.org/10.1103/PhysRevLett.130.206501} {\bibfield  {journal}
  {\bibinfo  {journal} {Phys. Rev. Lett.}\ }\textbf {\bibinfo {volume} {130}},\
  \bibinfo {pages} {206501} (\bibinfo {year} {2023})}\BibitemShut {NoStop}%
\bibitem [{\citenamefont {Lieb}\ and\ \citenamefont {Wu}(1972)}]{LiebWu}%
  \BibitemOpen
  \bibfield  {author} {\bibinfo {author} {\bibfnamefont {E.~H.}\ \bibnamefont
  {Lieb}}\ and\ \bibinfo {author} {\bibfnamefont {F.~Y.}\ \bibnamefont {Wu}},\
  }in\ \href@noop {} {\emph {\bibinfo {booktitle} {Phase transitions and
  critical phenomena: Exact Results}}},\ Vol.~\bibinfo {volume} {1},\ \bibinfo
  {editor} {edited by\ \bibinfo {editor} {\bibfnamefont {C.}~\bibnamefont
  {Domb}}\ and\ \bibinfo {editor} {\bibfnamefont {M.~S.}\ \bibnamefont
  {Green}}}\ (\bibinfo  {publisher} {Academic, London},\ \bibinfo {year}
  {1972})\ Chap.~\bibinfo {chapter} {8}, pp.\ \bibinfo {pages}
  {331--490}\BibitemShut {NoStop}%
\bibitem [{\citenamefont {Lieb}(1967{\natexlab{a}})}]{Lieb1967Residual}%
  \BibitemOpen
  \bibfield  {author} {\bibinfo {author} {\bibfnamefont {E.~H.}\ \bibnamefont
  {Lieb}},\ }\href {https://doi.org/10.1103/PhysRev.162.162} {\bibfield
  {journal} {\bibinfo  {journal} {Phys. Rev.}\ }\textbf {\bibinfo {volume}
  {162}},\ \bibinfo {pages} {162} (\bibinfo {year}
  {1967}{\natexlab{a}})}\BibitemShut {NoStop}%
\bibitem [{\citenamefont {Lieb}(1967{\natexlab{b}})}]{Lieb1967Exact}%
  \BibitemOpen
  \bibfield  {author} {\bibinfo {author} {\bibfnamefont {E.~H.}\ \bibnamefont
  {Lieb}},\ }\href {https://doi.org/10.1103/PhysRevLett.18.1046} {\bibfield
  {journal} {\bibinfo  {journal} {Phys. Rev. Lett.}\ }\textbf {\bibinfo
  {volume} {18}},\ \bibinfo {pages} {1046} (\bibinfo {year}
  {1967}{\natexlab{b}})}\BibitemShut {NoStop}%
\bibitem [{\citenamefont {Lieb}(1967{\natexlab{c}})}]{Lieb1967Exact2}%
  \BibitemOpen
  \bibfield  {author} {\bibinfo {author} {\bibfnamefont {E.~H.}\ \bibnamefont
  {Lieb}},\ }\href {https://doi.org/10.1103/PhysRevLett.19.108} {\bibfield
  {journal} {\bibinfo  {journal} {Phys. Rev. Lett.}\ }\textbf {\bibinfo
  {volume} {19}},\ \bibinfo {pages} {108} (\bibinfo {year}
  {1967}{\natexlab{c}})}\BibitemShut {NoStop}%
\bibitem [{\citenamefont {Sutherland}(1967)}]{Sutherland1967Exact}%
  \BibitemOpen
  \bibfield  {author} {\bibinfo {author} {\bibfnamefont {B.}~\bibnamefont
  {Sutherland}},\ }\href {https://doi.org/10.1103/PhysRevLett.19.103}
  {\bibfield  {journal} {\bibinfo  {journal} {Phys. Rev. Lett.}\ }\textbf
  {\bibinfo {volume} {19}},\ \bibinfo {pages} {103} (\bibinfo {year}
  {1967})}\BibitemShut {NoStop}%
\bibitem [{\citenamefont {Lieb}(1967{\natexlab{d}})}]{Lieb1967Exact3}%
  \BibitemOpen
  \bibfield  {author} {\bibinfo {author} {\bibfnamefont {E.~H.}\ \bibnamefont
  {Lieb}},\ }\href {https://doi.org/10.1103/PhysRevLett.18.692} {\bibfield
  {journal} {\bibinfo  {journal} {Phys. Rev. Lett.}\ }\textbf {\bibinfo
  {volume} {18}},\ \bibinfo {pages} {692} (\bibinfo {year}
  {1967}{\natexlab{d}})}\BibitemShut {NoStop}%
\bibitem [{\citenamefont {Pollock}\ and\ \citenamefont
  {Ceperley}(1987)}]{Pollock1987Path}%
  \BibitemOpen
  \bibfield  {author} {\bibinfo {author} {\bibfnamefont {E.~L.}\ \bibnamefont
  {Pollock}}\ and\ \bibinfo {author} {\bibfnamefont {D.~M.}\ \bibnamefont
  {Ceperley}},\ }\href {https://doi.org/10.1103/PhysRevB.36.8343} {\bibfield
  {journal} {\bibinfo  {journal} {Phys. Rev. B}\ }\textbf {\bibinfo {volume}
  {36}},\ \bibinfo {pages} {8343} (\bibinfo {year} {1987})}\BibitemShut
  {NoStop}%
\bibitem [{\citenamefont {Henelius}\ \emph {et~al.}(1998)\citenamefont
  {Henelius}, \citenamefont {Girvin},\ and\ \citenamefont
  {Sandvik}}]{Role1998Henelius}%
  \BibitemOpen
  \bibfield  {author} {\bibinfo {author} {\bibfnamefont {P.}~\bibnamefont
  {Henelius}}, \bibinfo {author} {\bibfnamefont {S.~M.}\ \bibnamefont
  {Girvin}},\ and\ \bibinfo {author} {\bibfnamefont {A.~W.}\ \bibnamefont
  {Sandvik}},\ }\href {https://doi.org/10.1103/PhysRevB.57.13382} {\bibfield
  {journal} {\bibinfo  {journal} {Phys. Rev. B}\ }\textbf {\bibinfo {volume}
  {57}},\ \bibinfo {pages} {13382} (\bibinfo {year} {1998})}\BibitemShut
  {NoStop}%
\bibitem [{\citenamefont {Leung}\ \emph {et~al.}(1996)\citenamefont {Leung},
  \citenamefont {Chiu},\ and\ \citenamefont {Runge}}]{leungcolumnar1996}%
  \BibitemOpen
  \bibfield  {author} {\bibinfo {author} {\bibfnamefont {P.~W.}\ \bibnamefont
  {Leung}}, \bibinfo {author} {\bibfnamefont {K.~C.}\ \bibnamefont {Chiu}},\
  and\ \bibinfo {author} {\bibfnamefont {K.~J.}\ \bibnamefont {Runge}},\ }\href
  {https://doi.org/10.1103/PhysRevB.54.12938} {\bibfield  {journal} {\bibinfo
  {journal} {Phys. Rev. B}\ }\textbf {\bibinfo {volume} {54}},\ \bibinfo
  {pages} {12938} (\bibinfo {year} {1996})}\BibitemShut {NoStop}%
\bibitem [{\citenamefont {Paiva}\ \emph {et~al.}(2004)\citenamefont {Paiva},
  \citenamefont {dos Santos}, \citenamefont {Scalettar},\ and\ \citenamefont
  {Denteneer}}]{paiva2004critical}%
  \BibitemOpen
  \bibfield  {author} {\bibinfo {author} {\bibfnamefont {T.}~\bibnamefont
  {Paiva}}, \bibinfo {author} {\bibfnamefont {R.~R.}\ \bibnamefont {dos
  Santos}}, \bibinfo {author} {\bibfnamefont {R.~T.}\ \bibnamefont
  {Scalettar}},\ and\ \bibinfo {author} {\bibfnamefont {P.~J.~H.}\ \bibnamefont
  {Denteneer}},\ }\href {https://doi.org/10.1103/PhysRevB.69.184501} {\bibfield
   {journal} {\bibinfo  {journal} {Phys. Rev. B}\ }\textbf {\bibinfo {volume}
  {69}},\ \bibinfo {pages} {184501} (\bibinfo {year} {2004})}\BibitemShut
  {NoStop}%
\bibitem [{\citenamefont {Chen}\ \emph {et~al.}(2021)\citenamefont {Chen},
  \citenamefont {Yuan}, \citenamefont {Qi},\ and\ \citenamefont
  {Meng}}]{chenFermi2021}%
  \BibitemOpen
  \bibfield  {author} {\bibinfo {author} {\bibfnamefont {C.}~\bibnamefont
  {Chen}}, \bibinfo {author} {\bibfnamefont {T.}~\bibnamefont {Yuan}}, \bibinfo
  {author} {\bibfnamefont {Y.}~\bibnamefont {Qi}},\ and\ \bibinfo {author}
  {\bibfnamefont {Z.~Y.}\ \bibnamefont {Meng}},\ }\href
  {https://doi.org/10.1103/PhysRevB.103.165131} {\bibfield  {journal} {\bibinfo
   {journal} {Phys. Rev. B}\ }\textbf {\bibinfo {volume} {103}},\ \bibinfo
  {pages} {165131} (\bibinfo {year} {2021})}\BibitemShut {NoStop}%
\bibitem [{\citenamefont {Costa}\ \emph {et~al.}(2018)\citenamefont {Costa},
  \citenamefont {Blommel}, \citenamefont {Chiu}, \citenamefont {Batrouni},\
  and\ \citenamefont {Scalettar}}]{costaPhonon2018}%
  \BibitemOpen
  \bibfield  {author} {\bibinfo {author} {\bibfnamefont {N.~C.}\ \bibnamefont
  {Costa}}, \bibinfo {author} {\bibfnamefont {T.}~\bibnamefont {Blommel}},
  \bibinfo {author} {\bibfnamefont {W.-T.}\ \bibnamefont {Chiu}}, \bibinfo
  {author} {\bibfnamefont {G.}~\bibnamefont {Batrouni}},\ and\ \bibinfo
  {author} {\bibfnamefont {R.~T.}\ \bibnamefont {Scalettar}},\ }\href
  {https://doi.org/10.1103/PhysRevLett.120.187003} {\bibfield  {journal}
  {\bibinfo  {journal} {Phys. Rev. Lett.}\ }\textbf {\bibinfo {volume} {120}},\
  \bibinfo {pages} {187003} (\bibinfo {year} {2018})}\BibitemShut {NoStop}%
\bibitem [{\citenamefont {Jiang}\ \emph {et~al.}(2022)\citenamefont {Jiang},
  \citenamefont {Liu}, \citenamefont {Klein}, \citenamefont {Wang},
  \citenamefont {Sun}, \citenamefont {Chubukov},\ and\ \citenamefont
  {Meng}}]{jiangMonte2022}%
  \BibitemOpen
  \bibfield  {author} {\bibinfo {author} {\bibfnamefont {W.}~\bibnamefont
  {Jiang}}, \bibinfo {author} {\bibfnamefont {Y.}~\bibnamefont {Liu}}, \bibinfo
  {author} {\bibfnamefont {A.}~\bibnamefont {Klein}}, \bibinfo {author}
  {\bibfnamefont {Y.}~\bibnamefont {Wang}}, \bibinfo {author} {\bibfnamefont
  {K.}~\bibnamefont {Sun}}, \bibinfo {author} {\bibfnamefont {A.~V.}\
  \bibnamefont {Chubukov}},\ and\ \bibinfo {author} {\bibfnamefont {Z.~Y.}\
  \bibnamefont {Meng}},\ }\href {https://doi.org/10.1038/s41467-022-30302-x}
  {\bibfield  {journal} {\bibinfo  {journal} {Nature Communications}\ }\textbf
  {\bibinfo {volume} {13}},\ \bibinfo {pages} {2655} (\bibinfo {year}
  {2022})}\BibitemShut {NoStop}%
\bibitem [{\citenamefont {Youngblood}\ \emph {et~al.}(1980)\citenamefont
  {Youngblood}, \citenamefont {Axe},\ and\ \citenamefont
  {McCoy}}]{younblood1980}%
  \BibitemOpen
  \bibfield  {author} {\bibinfo {author} {\bibfnamefont {R.}~\bibnamefont
  {Youngblood}}, \bibinfo {author} {\bibfnamefont {J.~D.}\ \bibnamefont
  {Axe}},\ and\ \bibinfo {author} {\bibfnamefont {B.~M.}\ \bibnamefont
  {McCoy}},\ }\href {https://doi.org/10.1103/PhysRevB.21.5212} {\bibfield
  {journal} {\bibinfo  {journal} {Phys. Rev. B}\ }\textbf {\bibinfo {volume}
  {21}},\ \bibinfo {pages} {5212} (\bibinfo {year} {1980})}\BibitemShut
  {NoStop}%
\bibitem [{\citenamefont {Sutherland}(1968)}]{sutherland68}%
  \BibitemOpen
  \bibfield  {author} {\bibinfo {author} {\bibfnamefont {B.}~\bibnamefont
  {Sutherland}},\ }\href
  {https://doi.org/https://doi.org/10.1016/0375-9601(68)90529-X} {\bibfield
  {journal} {\bibinfo  {journal} {Physics Letters A}\ }\textbf {\bibinfo
  {volume} {26}},\ \bibinfo {pages} {532} (\bibinfo {year} {1968})}\BibitemShut
  {NoStop}%
\bibitem [{\citenamefont {Falco}(2013)}]{falco2013}%
  \BibitemOpen
  \bibfield  {author} {\bibinfo {author} {\bibfnamefont {P.}~\bibnamefont
  {Falco}},\ }\href {https://doi.org/10.1103/PhysRevE.88.030103} {\bibfield
  {journal} {\bibinfo  {journal} {Phys. Rev. E}\ }\textbf {\bibinfo {volume}
  {88}},\ \bibinfo {pages} {030103} (\bibinfo {year} {2013})}\BibitemShut
  {NoStop}%
\bibitem [{\citenamefont {Fisher}\ and\ \citenamefont
  {Stephenson}(1963)}]{Fisher1963Statistical}%
  \BibitemOpen
  \bibfield  {author} {\bibinfo {author} {\bibfnamefont {M.~E.}\ \bibnamefont
  {Fisher}}\ and\ \bibinfo {author} {\bibfnamefont {J.}~\bibnamefont
  {Stephenson}},\ }\href {https://doi.org/10.1103/PhysRev.132.1411} {\bibfield
  {journal} {\bibinfo  {journal} {Phys. Rev.}\ }\textbf {\bibinfo {volume}
  {132}},\ \bibinfo {pages} {1411} (\bibinfo {year} {1963})}\BibitemShut
  {NoStop}%
\bibitem [{\citenamefont {Krauth}\ and\ \citenamefont
  {Moessner}(2003)}]{krauth2003pocket}%
  \BibitemOpen
  \bibfield  {author} {\bibinfo {author} {\bibfnamefont {W.}~\bibnamefont
  {Krauth}}\ and\ \bibinfo {author} {\bibfnamefont {R.}~\bibnamefont
  {Moessner}},\ }\href {https://doi.org/10.1103/PhysRevB.67.064503} {\bibfield
  {journal} {\bibinfo  {journal} {Phys. Rev. B}\ }\textbf {\bibinfo {volume}
  {67}},\ \bibinfo {pages} {064503} (\bibinfo {year} {2003})}\BibitemShut
  {NoStop}%
\bibitem [{\citenamefont {Andrews}\ \emph {et~al.}(1984)\citenamefont
  {Andrews}, \citenamefont {Baxter},\ and\ \citenamefont
  {Forrester}}]{andrews84}%
  \BibitemOpen
  \bibfield  {author} {\bibinfo {author} {\bibfnamefont {G.~E.}\ \bibnamefont
  {Andrews}}, \bibinfo {author} {\bibfnamefont {R.~J.}\ \bibnamefont
  {Baxter}},\ and\ \bibinfo {author} {\bibfnamefont {P.~J.}\ \bibnamefont
  {Forrester}},\ }\href {https://doi.org/10.1007/BF01014383} {\bibfield
  {journal} {\bibinfo  {journal} {J. Statist. Phys.}\ }\textbf {\bibinfo
  {volume} {35}},\ \bibinfo {pages} {193} (\bibinfo {year} {1984})}\BibitemShut
  {NoStop}%
\bibitem [{\citenamefont {Pasquier}(1987)}]{pasquier87}%
  \BibitemOpen
  \bibfield  {author} {\bibinfo {author} {\bibfnamefont {V.}~\bibnamefont
  {Pasquier}},\ }\href
  {https://doi.org/https://doi.org/10.1016/0550-3213(87)90332-4} {\bibfield
  {journal} {\bibinfo  {journal} {Nuclear Physics B}\ }\textbf {\bibinfo
  {volume} {285}},\ \bibinfo {pages} {162} (\bibinfo {year}
  {1987})}\BibitemShut {NoStop}%
\bibitem [{\citenamefont {Warnaar}\ \emph {et~al.}(1992)\citenamefont
  {Warnaar}, \citenamefont {Nienhuis},\ and\ \citenamefont
  {Seaton}}]{warnaar92}%
  \BibitemOpen
  \bibfield  {author} {\bibinfo {author} {\bibfnamefont {S.~O.}\ \bibnamefont
  {Warnaar}}, \bibinfo {author} {\bibfnamefont {B.}~\bibnamefont {Nienhuis}},\
  and\ \bibinfo {author} {\bibfnamefont {K.~A.}\ \bibnamefont {Seaton}},\
  }\href {https://doi.org/10.1103/PhysRevLett.69.710} {\bibfield  {journal}
  {\bibinfo  {journal} {Phys. Rev. Lett.}\ }\textbf {\bibinfo {volume} {69}},\
  \bibinfo {pages} {710} (\bibinfo {year} {1992})}\BibitemShut {NoStop}%
\bibitem [{\citenamefont {Bl\"ote}\ and\ \citenamefont
  {Nightingale}(1993)}]{blote93}%
  \BibitemOpen
  \bibfield  {author} {\bibinfo {author} {\bibfnamefont {H.~W.~J.}\
  \bibnamefont {Bl\"ote}}\ and\ \bibinfo {author} {\bibfnamefont {M.~P.}\
  \bibnamefont {Nightingale}},\ }\href
  {https://doi.org/10.1103/PhysRevB.47.15046} {\bibfield  {journal} {\bibinfo
  {journal} {Phys. Rev. B}\ }\textbf {\bibinfo {volume} {47}},\ \bibinfo
  {pages} {15046} (\bibinfo {year} {1993})}\BibitemShut {NoStop}%
\bibitem [{\citenamefont {Bl\"ote}\ and\ \citenamefont
  {Nienhuis}(1994)}]{blote94}%
  \BibitemOpen
  \bibfield  {author} {\bibinfo {author} {\bibfnamefont {H.~W.~J.}\
  \bibnamefont {Bl\"ote}}\ and\ \bibinfo {author} {\bibfnamefont
  {B.}~\bibnamefont {Nienhuis}},\ }\href
  {https://doi.org/10.1103/PhysRevLett.72.1372} {\bibfield  {journal} {\bibinfo
   {journal} {Phys. Rev. Lett.}\ }\textbf {\bibinfo {volume} {72}},\ \bibinfo
  {pages} {1372} (\bibinfo {year} {1994})}\BibitemShut {NoStop}%
\bibitem [{\citenamefont {Kondev}\ and\ \citenamefont
  {Henley}(1995)}]{kondev96}%
  \BibitemOpen
  \bibfield  {author} {\bibinfo {author} {\bibfnamefont {J.}~\bibnamefont
  {Kondev}}\ and\ \bibinfo {author} {\bibfnamefont {C.~L.}\ \bibnamefont
  {Henley}},\ }\href {https://doi.org/10.1103/PhysRevB.52.6628} {\bibfield
  {journal} {\bibinfo  {journal} {Phys. Rev. B}\ }\textbf {\bibinfo {volume}
  {52}},\ \bibinfo {pages} {6628} (\bibinfo {year} {1995})}\BibitemShut
  {NoStop}%
\bibitem [{\citenamefont {Kondev}\ and\ \citenamefont
  {Henley}(1994)}]{kondev94}%
  \BibitemOpen
  \bibfield  {author} {\bibinfo {author} {\bibfnamefont {J.}~\bibnamefont
  {Kondev}}\ and\ \bibinfo {author} {\bibfnamefont {C.~L.}\ \bibnamefont
  {Henley}},\ }\href {https://doi.org/10.1103/PhysRevLett.73.2786} {\bibfield
  {journal} {\bibinfo  {journal} {Phys. Rev. Lett.}\ }\textbf {\bibinfo
  {volume} {73}},\ \bibinfo {pages} {2786} (\bibinfo {year}
  {1994})}\BibitemShut {NoStop}%
\bibitem [{\citenamefont {Wilkins}\ and\ \citenamefont
  {Powell}(2020)}]{wilkins2020}%
  \BibitemOpen
  \bibfield  {author} {\bibinfo {author} {\bibfnamefont {N.}~\bibnamefont
  {Wilkins}}\ and\ \bibinfo {author} {\bibfnamefont {S.}~\bibnamefont
  {Powell}},\ }\href {https://doi.org/10.1103/PhysRevB.102.174431} {\bibfield
  {journal} {\bibinfo  {journal} {Phys. Rev. B}\ }\textbf {\bibinfo {volume}
  {102}},\ \bibinfo {pages} {174431} (\bibinfo {year} {2020})}\BibitemShut
  {NoStop}%
\bibitem [{\citenamefont {S{\'e}n{\'e}chal}(2004)}]{senechal2004introduction}%
  \BibitemOpen
  \bibfield  {author} {\bibinfo {author} {\bibfnamefont {D.}~\bibnamefont
  {S{\'e}n{\'e}chal}},\ }in\ \href@noop {} {\emph {\bibinfo {booktitle}
  {Theoretical Methods for Strongly Correlated Electrons}}}\ (\bibinfo
  {publisher} {Springer},\ \bibinfo {year} {2004})\ pp.\ \bibinfo {pages}
  {139--186}\BibitemShut {NoStop}%
\bibitem [{\citenamefont {Lukyanov}\ and\ \citenamefont
  {Terras}(2003)}]{lukyanov_long-distance_2003}%
  \BibitemOpen
  \bibfield  {author} {\bibinfo {author} {\bibfnamefont {S.}~\bibnamefont
  {Lukyanov}}\ and\ \bibinfo {author} {\bibfnamefont {V.}~\bibnamefont
  {Terras}},\ }\href {https://doi.org/10.1016/S0550-3213(02)01141-0} {\bibfield
   {journal} {\bibinfo  {journal} {Nuclear Physics B}\ }\textbf {\bibinfo
  {volume} {654}},\ \bibinfo {pages} {323} (\bibinfo {year}
  {2003})}\BibitemShut {NoStop}%
\bibitem [{\citenamefont {Jos\'e}\ \emph {et~al.}(1977)\citenamefont {Jos\'e},
  \citenamefont {Kadanoff}, \citenamefont {Kirkpatrick},\ and\ \citenamefont
  {Nelson}}]{jose77}%
  \BibitemOpen
  \bibfield  {author} {\bibinfo {author} {\bibfnamefont {J.~V.}\ \bibnamefont
  {Jos\'e}}, \bibinfo {author} {\bibfnamefont {L.~P.}\ \bibnamefont
  {Kadanoff}}, \bibinfo {author} {\bibfnamefont {S.}~\bibnamefont
  {Kirkpatrick}},\ and\ \bibinfo {author} {\bibfnamefont {D.~R.}\ \bibnamefont
  {Nelson}},\ }\href {https://doi.org/10.1103/PhysRevB.16.1217} {\bibfield
  {journal} {\bibinfo  {journal} {Phys. Rev. B}\ }\textbf {\bibinfo {volume}
  {16}},\ \bibinfo {pages} {1217} (\bibinfo {year} {1977})}\BibitemShut
  {NoStop}%
\bibitem [{\citenamefont {Amit}\ \emph {et~al.}(1980)\citenamefont {Amit},
  \citenamefont {Goldschmidt},\ and\ \citenamefont {Grinstein}}]{amit80}%
  \BibitemOpen
  \bibfield  {author} {\bibinfo {author} {\bibfnamefont {D.~J.}\ \bibnamefont
  {Amit}}, \bibinfo {author} {\bibfnamefont {Y.~Y.}\ \bibnamefont
  {Goldschmidt}},\ and\ \bibinfo {author} {\bibfnamefont {S.}~\bibnamefont
  {Grinstein}},\ }\href {https://doi.org/10.1088/0305-4470/13/2/024} {\bibfield
   {journal} {\bibinfo  {journal} {Journal of Physics A: Mathematical and
  General}\ }\textbf {\bibinfo {volume} {13}},\ \bibinfo {pages} {585}
  (\bibinfo {year} {1980})}\BibitemShut {NoStop}%
\bibitem [{Note1()}]{Note1}%
  \BibitemOpen
  \bibinfo {note} {The ice-point corresponds to the XXZ spin chain at $\Delta =
  1/2$ in notations where $\Delta >0$ corresponds to ferromagnetic
  interactions. The value of $g$ is given by $\Delta =\protect \qopname \relax
  o{cos}(\pi g)$, see e.g. Ref.~\cite
  {lukyanov_long-distance_2003,hikihara1998correlation}.}\BibitemShut {Stop}%
\bibitem [{\citenamefont {Kosterlitz}\ and\ \citenamefont
  {Thouless}(1973)}]{KT73}%
  \BibitemOpen
  \bibfield  {author} {\bibinfo {author} {\bibfnamefont {J.~M.}\ \bibnamefont
  {Kosterlitz}}\ and\ \bibinfo {author} {\bibfnamefont {D.~J.}\ \bibnamefont
  {Thouless}},\ }\href {https://doi.org/10.1088/0022-3719/6/7/010} {\bibfield
  {journal} {\bibinfo  {journal} {Journal of Physics C: Solid State Physics}\
  }\textbf {\bibinfo {volume} {6}},\ \bibinfo {pages} {1181} (\bibinfo {year}
  {1973})}\BibitemShut {NoStop}%
\bibitem [{\citenamefont {Kosterlitz}(1974)}]{Kosterlitz74}%
  \BibitemOpen
  \bibfield  {author} {\bibinfo {author} {\bibfnamefont {J.~M.}\ \bibnamefont
  {Kosterlitz}},\ }\href {https://doi.org/10.1088/0022-3719/7/6/005} {\bibfield
   {journal} {\bibinfo  {journal} {Journal of Physics C: Solid State Physics}\
  }\textbf {\bibinfo {volume} {7}},\ \bibinfo {pages} {1046} (\bibinfo {year}
  {1974})}\BibitemShut {NoStop}%
\bibitem [{\citenamefont {Ginsparg}(1988)}]{ginsparg1988curiosities}%
  \BibitemOpen
  \bibfield  {author} {\bibinfo {author} {\bibfnamefont {P.}~\bibnamefont
  {Ginsparg}},\ }\href@noop {} {\bibfield  {journal} {\bibinfo  {journal}
  {Nuclear Physics B}\ }\textbf {\bibinfo {volume} {295}},\ \bibinfo {pages}
  {153} (\bibinfo {year} {1988})}\BibitemShut {NoStop}%
\bibitem [{\citenamefont {Archambault}\ \emph {et~al.}(1998)\citenamefont
  {Archambault}, \citenamefont {Bramwell}, \citenamefont {Fortin},
  \citenamefont {Holdsworth}, \citenamefont {Peysson},\ and\ \citenamefont
  {Pinton}}]{archambaultUniversal1998}%
  \BibitemOpen
  \bibfield  {author} {\bibinfo {author} {\bibfnamefont {P.}~\bibnamefont
  {Archambault}}, \bibinfo {author} {\bibfnamefont {S.~T.}\ \bibnamefont
  {Bramwell}}, \bibinfo {author} {\bibfnamefont {J.-Y.}\ \bibnamefont
  {Fortin}}, \bibinfo {author} {\bibfnamefont {P.~C.~W.}\ \bibnamefont
  {Holdsworth}}, \bibinfo {author} {\bibfnamefont {S.}~\bibnamefont
  {Peysson}},\ and\ \bibinfo {author} {\bibfnamefont {J.-F.}\ \bibnamefont
  {Pinton}},\ }\href {https://doi.org/10.1063/1.367855} {\bibfield  {journal}
  {\bibinfo  {journal} {Journal of Applied Physics}\ }\textbf {\bibinfo
  {volume} {83}},\ \bibinfo {pages} {7234} (\bibinfo {year} {1998})},\ \Eprint
  {https://arxiv.org/abs/https://doi.org/10.1063/1.367855}
  {https://doi.org/10.1063/1.367855} \BibitemShut {NoStop}%
\bibitem [{\citenamefont {Atchison}\ \emph {et~al.}(2019)\citenamefont
  {Atchison}, \citenamefont {Bhullar}, \citenamefont {Norman},\ and\
  \citenamefont {Venus}}]{atchisonFinite2019}%
  \BibitemOpen
  \bibfield  {author} {\bibinfo {author} {\bibfnamefont {J.}~\bibnamefont
  {Atchison}}, \bibinfo {author} {\bibfnamefont {A.}~\bibnamefont {Bhullar}},
  \bibinfo {author} {\bibfnamefont {B.}~\bibnamefont {Norman}},\ and\ \bibinfo
  {author} {\bibfnamefont {D.}~\bibnamefont {Venus}},\ }\href
  {https://doi.org/10.1103/PhysRevB.99.125425} {\bibfield  {journal} {\bibinfo
  {journal} {Phys. Rev. B}\ }\textbf {\bibinfo {volume} {99}},\ \bibinfo
  {pages} {125425} (\bibinfo {year} {2019})}\BibitemShut {NoStop}%
\bibitem [{\citenamefont {Isakov}\ and\ \citenamefont
  {Moessner}(2003)}]{isakov2003interplay}%
  \BibitemOpen
  \bibfield  {author} {\bibinfo {author} {\bibfnamefont {S.~V.}\ \bibnamefont
  {Isakov}}\ and\ \bibinfo {author} {\bibfnamefont {R.}~\bibnamefont
  {Moessner}},\ }\href {https://doi.org/10.1103/PhysRevB.68.104409} {\bibfield
  {journal} {\bibinfo  {journal} {Phys. Rev. B}\ }\textbf {\bibinfo {volume}
  {68}},\ \bibinfo {pages} {104409} (\bibinfo {year} {2003})}\BibitemShut
  {NoStop}%
\bibitem [{\citenamefont {Hikihara}\ and\ \citenamefont
  {Furusaki}(1998)}]{hikihara1998correlation}%
  \BibitemOpen
  \bibfield  {author} {\bibinfo {author} {\bibfnamefont {T.}~\bibnamefont
  {Hikihara}}\ and\ \bibinfo {author} {\bibfnamefont {A.}~\bibnamefont
  {Furusaki}},\ }\href
  {https://journals.aps.org/prb/abstract/10.1103/PhysRevB.58.R583} {\bibfield
  {journal} {\bibinfo  {journal} {Physical Review B}\ }\textbf {\bibinfo
  {volume} {58}},\ \bibinfo {pages} {R583} (\bibinfo {year}
  {1998})}\BibitemShut {NoStop}%
\bibitem [{\citenamefont {Castelnovo}\ \emph {et~al.}(2006)\citenamefont
  {Castelnovo}, \citenamefont {Chamon}, \citenamefont {Mudry},\ and\
  \citenamefont {Pujol}}]{castelnovoHigh2006}%
  \BibitemOpen
  \bibfield  {author} {\bibinfo {author} {\bibfnamefont {C.}~\bibnamefont
  {Castelnovo}}, \bibinfo {author} {\bibfnamefont {C.}~\bibnamefont {Chamon}},
  \bibinfo {author} {\bibfnamefont {C.}~\bibnamefont {Mudry}},\ and\ \bibinfo
  {author} {\bibfnamefont {P.}~\bibnamefont {Pujol}},\ }\href
  {https://doi.org/10.1103/PhysRevB.73.144411} {\bibfield  {journal} {\bibinfo
  {journal} {Phys. Rev. B}\ }\textbf {\bibinfo {volume} {73}},\ \bibinfo
  {pages} {144411} (\bibinfo {year} {2006})}\BibitemShut {NoStop}%
\bibitem [{\citenamefont {Yan}\ \emph {et~al.}(2019)\citenamefont {Yan},
  \citenamefont {Wu}, \citenamefont {Liu}, \citenamefont {Sylju\aa{}sen},
  \citenamefont {Lou},\ and\ \citenamefont {Chen}}]{yanSweeping2019}%
  \BibitemOpen
  \bibfield  {author} {\bibinfo {author} {\bibfnamefont {Z.}~\bibnamefont
  {Yan}}, \bibinfo {author} {\bibfnamefont {Y.}~\bibnamefont {Wu}}, \bibinfo
  {author} {\bibfnamefont {C.}~\bibnamefont {Liu}}, \bibinfo {author}
  {\bibfnamefont {O.~F.}\ \bibnamefont {Sylju\aa{}sen}}, \bibinfo {author}
  {\bibfnamefont {J.}~\bibnamefont {Lou}},\ and\ \bibinfo {author}
  {\bibfnamefont {Y.}~\bibnamefont {Chen}},\ }\href
  {https://doi.org/10.1103/PhysRevB.99.165135} {\bibfield  {journal} {\bibinfo
  {journal} {Phys. Rev. B}\ }\textbf {\bibinfo {volume} {99}},\ \bibinfo
  {pages} {165135} (\bibinfo {year} {2019})}\BibitemShut {NoStop}%
\bibitem [{\citenamefont {Yan}(2022)}]{yanGlobal2022}%
  \BibitemOpen
  \bibfield  {author} {\bibinfo {author} {\bibfnamefont {Z.}~\bibnamefont
  {Yan}},\ }\href {https://doi.org/10.1103/PhysRevB.105.184432} {\bibfield
  {journal} {\bibinfo  {journal} {Phys. Rev. B}\ }\textbf {\bibinfo {volume}
  {105}},\ \bibinfo {pages} {184432} (\bibinfo {year} {2022})}\BibitemShut
  {NoStop}%
\bibitem [{\citenamefont {Scholl}\ \emph {et~al.}(2021)\citenamefont {Scholl},
  \citenamefont {Schuler}, \citenamefont {Williams}, \citenamefont
  {Eberharter}, \citenamefont {Barredo}, \citenamefont {Schymik}, \citenamefont
  {Lienhard}, \citenamefont {Henry}, \citenamefont {Lang}, \citenamefont
  {Lahaye}, \citenamefont {Läuchli},\ and\ \citenamefont
  {Browaeys}}]{schollQuantum2021}%
  \BibitemOpen
  \bibfield  {author} {\bibinfo {author} {\bibfnamefont {P.}~\bibnamefont
  {Scholl}}, \bibinfo {author} {\bibfnamefont {M.}~\bibnamefont {Schuler}},
  \bibinfo {author} {\bibfnamefont {H.~J.}\ \bibnamefont {Williams}}, \bibinfo
  {author} {\bibfnamefont {A.~A.}\ \bibnamefont {Eberharter}}, \bibinfo
  {author} {\bibfnamefont {D.}~\bibnamefont {Barredo}}, \bibinfo {author}
  {\bibfnamefont {K.-N.}\ \bibnamefont {Schymik}}, \bibinfo {author}
  {\bibfnamefont {V.}~\bibnamefont {Lienhard}}, \bibinfo {author}
  {\bibfnamefont {L.-P.}\ \bibnamefont {Henry}}, \bibinfo {author}
  {\bibfnamefont {T.~C.}\ \bibnamefont {Lang}}, \bibinfo {author}
  {\bibfnamefont {T.}~\bibnamefont {Lahaye}}, \bibinfo {author} {\bibfnamefont
  {A.~M.}\ \bibnamefont {Läuchli}},\ and\ \bibinfo {author} {\bibfnamefont
  {A.}~\bibnamefont {Browaeys}},\ }\href
  {https://doi.org/10.1038/s41586-021-03585-1} {\bibfield  {journal} {\bibinfo
  {journal} {Nature}\ }\textbf {\bibinfo {volume} {595}},\ \bibinfo {pages}
  {233 } (\bibinfo {year} {2021})}\BibitemShut {NoStop}%
\end{thebibliography}%

\setcounter{equation}{0}
\setcounter{figure}{0}
\renewcommand{\theequation}{S\arabic{equation}}
\renewcommand{\thefigure}{S\arabic{figure}}

\clearpage
\newpage
\setcounter{page}{1}
\appendix	

\section{Directed loop algorithm for loop models}
\label{sec:appI}
\begin{figure}[h]
\includegraphics[width=\columnwidth]{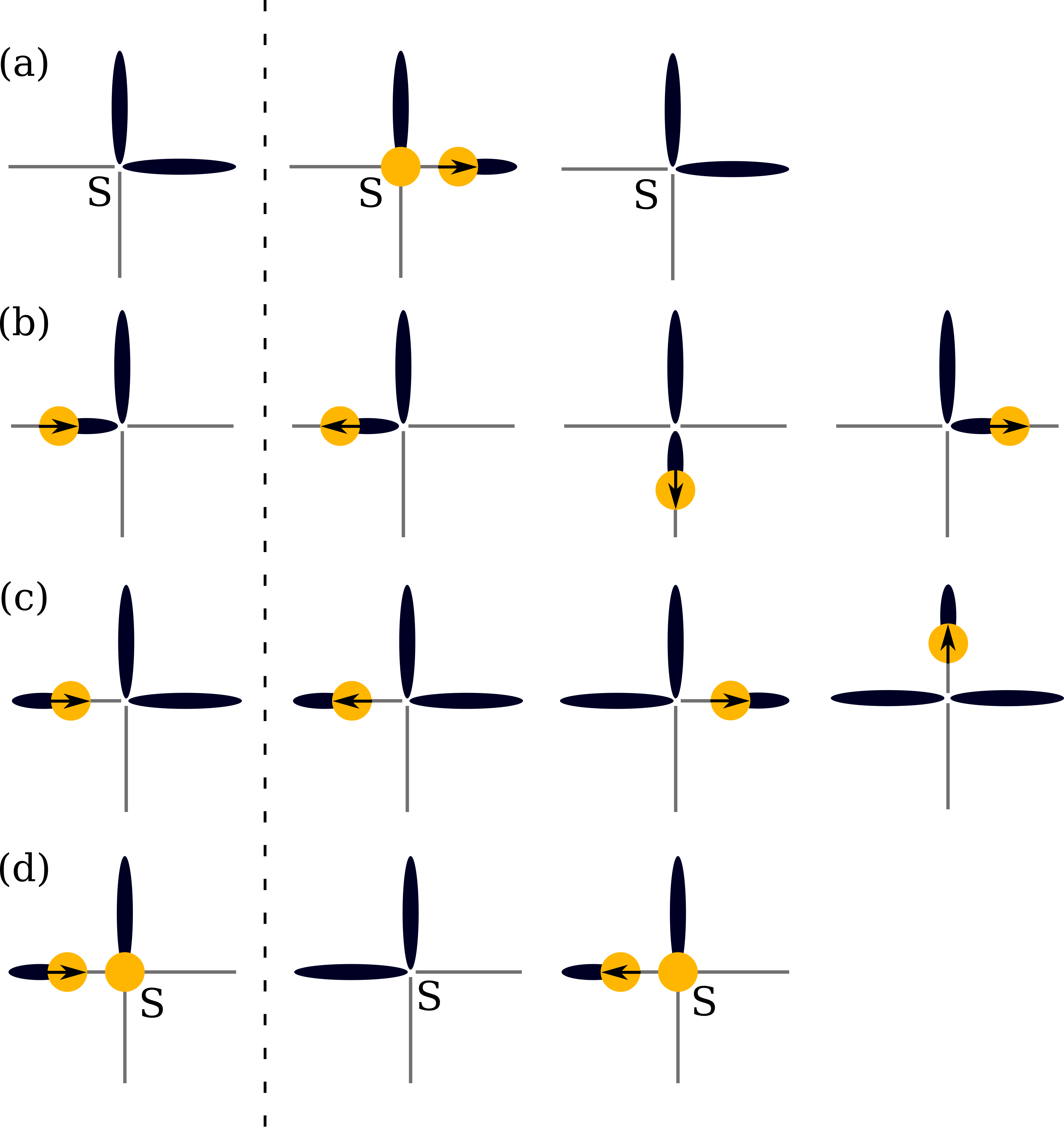}
\caption{
Directed loop Monte Carlo update steps for the fully packed loop model. The local configuration to the left of the dotted line can transition into one of the configurations to its right with probabilities determined by detailed balance.
\label{fig:directedLoop}}
\end{figure}
We estimate the thermal averages of observables in the classical loop model as averages over Monte Carlo samples generated by a directed loop algorithm~\cite{barkema1998monte,alet2006classical,sandvik2006correlations,syljuasenquantum2002,syljuaasen2004directed,aletdirected2003} tailored for the loop model. The algorithm is summarized below.

1. Given a fully packed loop configuration $\mathcal{C}$, we pick with uniform probability a site $S$; and then choose one of the two occupied edges around $S$. With a Metropolis probability $p_0={\rm min}(1,e^{\beta (E(\mathcal{C})-E(\mathcal{C}'))})$, the dimer on this edge is replaced by half a dimer (Fig. \ref{fig:directedLoop}a). The new configuration $\mathcal{C}'$ has monomers on the site $S$ and at the end (monomer $M$) of the dimer. The monomer $M$ has a binary valued `momentum' internal degree of freedom that is, initially,  directed into the dimer and away from $S$. In calculating the configurations $E$, it is assumed that the interaction between half-dimers and parallel dimers is half that of full dimers. Note that parallel dimers in a plaquette interact only if the other two edges are empty.
With probability $1-p_0$ the move is abandoned in this first step itself.

2. If the monomer $M$ is moving into a dimer, annihilating it in the process (Fig. \ref{fig:directedLoop}b), it can, subsequently, create a dimer on one of the two previously empty edges connected to the site ahead or the monomer can just reverse its direction. Transition probabilities are chosen to satisfy detailed balance as described further below. 

3. If the monomer is moving away from the dimer, growing a dimer in the process (Fig. \ref{fig:directedLoop}c), it can, subsequently, destroy one of the two dimers connected to the node ahead or the monomer can just reverse its direction. Transition probabilities are chosen to satisfy detailed balance as described further below. 

4. We repeat steps (2) and (3) till the loop closes. 
If in the current configuration $\mathcal{C}'$, the monomer $M$ sees the starting site $S$ ahead of it (Fig. \ref{fig:directedLoop}d), the loop can terminate and produce a fully packed configuration $\mathcal{C}$ with a Metropolis probability $p_{\rm term}={\rm min}(1,e^{\beta(E(\mathcal{C}')-E(\mathcal{C}))})$. With probability $1-p_{\rm term}$, the $M$ reverses the direction instead.

The probabilities $p$ in steps (2) and (3) are chosen to satisfy detailed balance. As shown in Fig. \ref{fig:directedLoop}b,c the current configuration $\mathcal{C}$ can transition into $\mathcal{C}'_0$ with the monomer direction reversed or two other configurations $\mathcal{C}'_{1,2}$. The probability of transition to $\mathcal{C}'_{1,2}$ is given by 
\begin{equation}
\frac{W_{1,2}}{Z-\min(W_1,W_2,W_0)}
\end{equation}
where $W_i=e^{-\beta E(\mathcal{C}_i)}$, $Z=\sum_i W_i$.

The monomer correlator $M(r)$ is computed as the histogram of displacements between $S$ and $M$. In the description of the algorithm above we have used the convention that the monomer sits at the center of the edge. We can instead choose to place the monomer $M$ at some position $x\in[0,1]$ on the dimer and associate an interaction energy between parallel full-dimers and the `partial' dimers that is commensurate with $x$. We find that the choice of $x$ adds a short range correction that does not affect the scaling properties.

\begin{figure}[htp!]
	\centering
	\includegraphics[width=1\columnwidth]{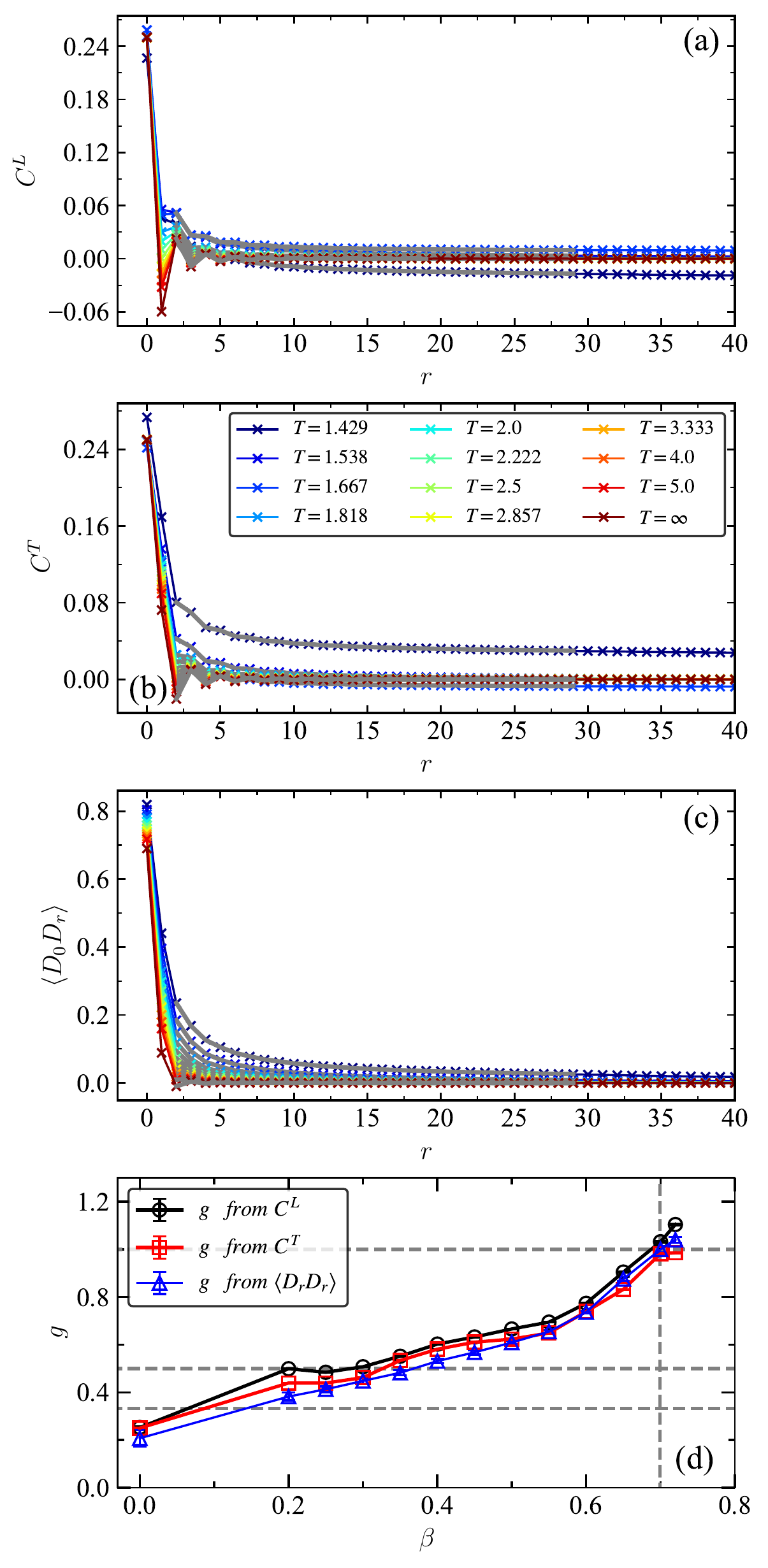}
	\caption{The equal-time loop-segment correlation functions of (a) longitudinal loop-segment correlations $C^L$, (b) transverse correlations $C^T$, and (c) $\langle D_0D_r \rangle$ in the Eqs.~\eqref{eq:eq5}, ~\eqref{eq:eq6}, and ~\eqref{eq:eq8} in the main text. The system size here is $L=256$. (d) The Coulomb gas constant $g$, as extracted from the various fits presented in previous panels,  as a function of inverse temperature $\beta$.} 
	\label{fig:S2}
\end{figure}

\section{Raw data of the correlation functions}
\label{sec:appII}

We present in Fig.~\ref{fig:S2} (a) and (b) the correlators $C^{L}(r)$ and $C^{T}(r)$ (measured in the ${\bf r}=(r,0)$ direction) as well as the fits to the expressions Eq.~\eqref{eq:eq5} and Eq.~\eqref{eq:eq6}.
We furthermore present the correlator $\langle D_0 D_r \rangle$ associated to the order parameter in  Fig.~\ref{fig:S2} (c), which we fit to a single power-law as its leading contribution should decay as $r^{-1/g}$. From these fits, we obtain estimates of $g$ represented in the panel Fig.~\ref{fig:S2} (d), which are in overall agreement to those obtained from fits to the adapted linear combinations of correlators presented in the main text (see Fig.~\ref{fig:5}), albeit with slightly larger fluctuations at high temperature.

\begin{figure}[!h]
	\centering
	\includegraphics[width=1\columnwidth]{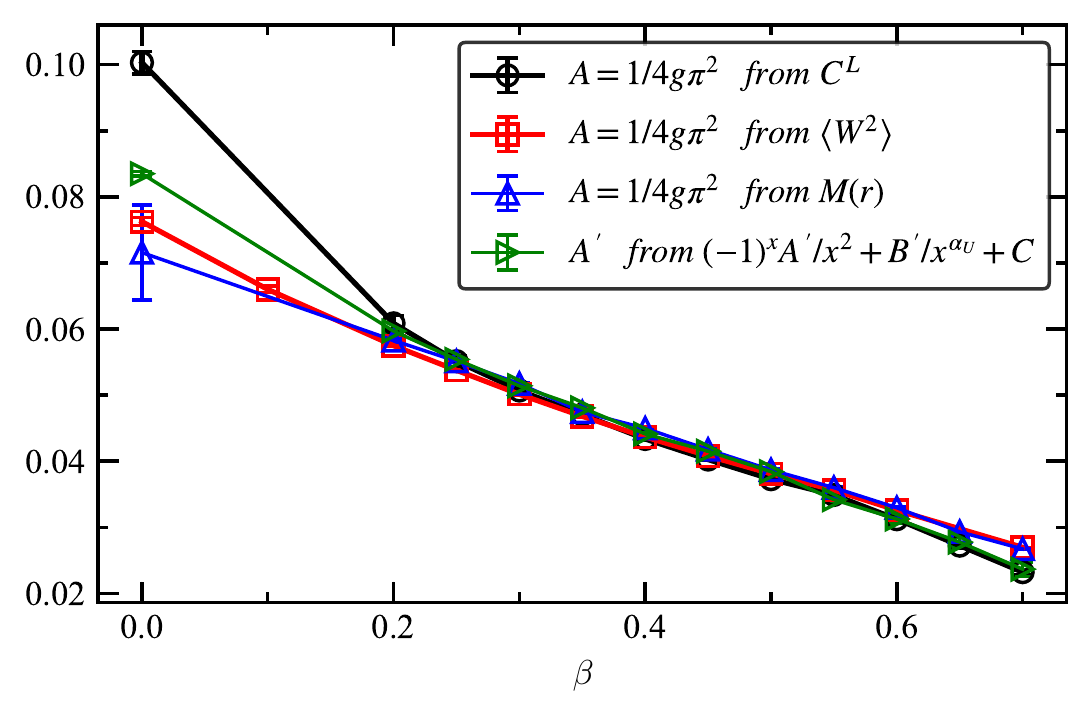}
	\caption{ Fits to obtain the coefficient $A$ in Eq.~\eqref{eq:eq12} for the longitudinal loop-segment correlations $C^L$. Here we show the estimated values of $A$ from $1/4g\pi^2$, where $g$ is obtained in three ways -  (black circular markers) from fitting to $(-1)^r A'/r^{\alpha_S} + B'/r^{1/g} + C$, (red squares) from $\langle W^2\rangle$ and (blue triangles $\Delta$) from $M(r)$ (last two as described in Fig.~\ref{fig:5}). We also show $A'$ obtained directly from the fit to $(-1)^r A'/r^{2} + B'/r^{\alpha_U} + C$. The four curves are consistent when $\beta$ is large. } 
	\label{fig:S3}
\end{figure}

Aside from  the evaluation of the Coulomb gas constant, we also comment in this Appendix on the evaluation of the amplitude [denoted $A$ in Eq.~\eqref{eq:eq5},~\eqref{eq:eq6} and~\eqref{eq:eq7}] of the staggered part of loop segment correlators, which is also expected to be universal (with $g$). We evaluate this amplitude from measurement of the longitudinal correlator $C^L(r)$ in four different ways (see Fig.~\ref{fig:S3}): (i) assuming $A=1/(4 \pi^2 g)$ where $g$ is obtained from the scaling $(-1)^r A'/r^{\alpha_S} + B'/r^{1/g} + C$. (ii) and (iii) are $A=1/(4 \pi^2 g)$ where $g$ is from the winding number fluctuation $\langle W^2\rangle$ and the monomer correlator $M(r)$ respectively. (iv) $A'$ obtained from the scaling form $(-1)^r A'/r^{2} + B'/r^{\alpha_U} + C$.

We note an overall good agreement between all determinations of this amplitude, as soon as $\beta\geq 0.2$, albeit with some small discrepancy at $\beta=0$. Overall these data are consistent with the prediction $A=1/(4 \pi^2 g)$.

\end{document}